 \documentclass[traditabstract]{aa}
\usepackage{txfonts}
\usepackage{graphicx}
\usepackage{natbib}
\bibpunct{(}{)}{;}{a}{}{,}
\newcommand{\Nabla}{\vec{\nabla}}
\newcommand{\vB}{\vec{B}}

\newcommand{\vJ}{\vec{J}}
\newcommand{\vv}{\vec{v}}
\newcommand{\vx}{\vec{x}}
\newcommand{\divb}{\Nabla \cdot \vB}
\newcommand{\sj}{\sigma_J}
\newcommand{\eg}{e.g., } 
\begin{document}

\title{Testing magnetofrictional extrapolation with the 
       Titov-D\'emoulin model of solar active regions}

\author{G. Valori \inst{1} \and  B. Kliem \inst{2,3,4} \and T. T\"or\"ok \inst{1} \and V. S. Titov \inst{5}}
\institute{
 {LESIA, Observatoire de Paris, CNRS, UPMC, Universit\'e Paris Diderot, 5 place Jules Janssen, 92190 Meudon, France}
\and 
{Universit{\"a}t Potsdam, Institut f{\"u}r Physik und Astronomie, 14482 Potsdam, Germany}
\and 
{University College London, Mullard Space Science Laboratory, Holmbury St.~Mary, Dorking, Surrey, RH5 6NT, UK}
\and 
{Naval Research Laboratory, Space Science Division, Washington, DC 20375, USA}
\and 
{Predictive Science, Inc., 9990 Mesa Rim Road, Suite 170, San Diego, CA 92121-2910, USA}
}
 \date{Received 15 March 2010 / Accepted 28 April 2010}

\abstract{
We examine the nonlinear magnetofrictional extrapolation scheme using the solar active region model by Titov and D\'emoulin as test field. 
This model consists of an arched, line-tied current channel held in force-free equilibrium by the potential field of a bipolar flux distribution in the bottom boundary. 
A modified version, having a parabolic current density profile, is employed here.
We find that the equilibrium is reconstructed with very high accuracy in a representative range of parameter space, using only the vector field in the bottom boundary as input. 
Structural features formed in the interface between the flux rope and the surrounding arcade---``hyperbolic flux tube'' and ``bald patch separatrix surface''---are reliably reproduced, as are the flux rope twist and the energy and helicity of the configuration. 
This demonstrates that force-free fields containing these basic structural elements of solar active regions can be obtained by extrapolation. 
The influence of the chosen initial condition on the accuracy of reconstruction is also addressed, confirming that the initial field that best matches the external potential field of the model quite naturally leads to the best reconstruction. 
Extrapolating the magnetogram of a Titov-D\'emoulin equilibrium in the unstable range of parameter space yields a sequence of two opposing evolutionary phases which clearly indicate the unstable nature of the configuration: a partial buildup of the flux rope with rising free energy is followed by destruction of the rope, losing most of the free energy. 
}
\maketitle
%
\section{Introduction}\label{s:intro}

The computation of the coronal magnetic field from boundary data is the only available technique to obtain fully three-dimensional information, i.e., the complete field vector in the whole coronal volume bounded below by the magnetogram. 
Deductions of the field from spectral lines formed in the coronal plasma, from radio maps across a range of microwave frequencies, or from observations of specific structures such as prominences, are restricted to two-dimensional projections or to cuts through the volume, or do not yield the full vector. 
The three-dimensional information is required to advance a wide array of issues in coronal physics, for example the structure of prominences, the onset of eruptions, and how the coronal field is connected to the Sun's interior and the solar wind.

In equilibrium, the coronal magnetic field is nearly force-free in the major part of active regions, except their upper periphery \cite[see, \eg][]{1995ApJ...439..474M, 2002ApJ...568..422M, 2001SoPh..203...71G}. 
Hence, to a very good approximation it can be described by
\begin{equation}
\Nabla \times \vB = \alpha \vB \qquad \mbox{with} \qquad \Nabla \cdot \vB=0.
\label{eq:fff}
\end{equation}
Here $\alpha(\vx)$ is a scalar function that is constant along each individual field line (which follows directly from Eqs.~[\ref{eq:fff}]) but in general has different values on different field lines.

The computation of force-free coronal fields can be formulated as a boundary-value problem for Eqs.~(\ref{eq:fff}). 
Such extrapolation can be performed at different levels of approximation for the field. 
If $\alpha$ is assumed to be constant, then the first of Eqs.~(\ref{eq:fff}) is linear, and only the knowledge of the normal field component is needed as boundary condition for the extrapolation.
The solution of the linear problem was given in closed form in \cite{1972SoPh...25..127N, 1977ApJ...212..873C, 1978SoPh...58..215S}. 
Linear methods have several intrinsic limitations, the most severe one being that they cannot match most observed magnetograms closely because $\alpha$ is often found to vary strongly across active regions \citep[\eg][]{2002A&A...392.1119R}.
The particular case of $\alpha=0$ yields a potential field, which is inadequate for many purposes, since it does not contain any free energy to power coronal activity.
Therefore, it is necessary to proceed beyond the linear approximation and to solve the nonlinear extrapolation problem, permitting $\alpha(\vx)$ to vary across the field \citep{1981SoPh...69..343S,1997SoPh..174..191M,1999A&A...350.1051A}. 
Due to the nonlinearity, such extrapolation requires a numerical approach. 
Various numerical schemes and implementations have been developed to construct a solution; for a recent review see \citet{2006SoPh..235..161S}.

A partially alternative strategy, the flux rope insertion method \citep{2004ApJ...612..519V, ballegooijen07}, permits to model structures that are supposed to contain a flux rope, particularly filament channels. 
Here an extrapolated potential field is replaced by a flux rope in part of the volume, and the resulting configuration is numerically relaxed to find an equilibrium. 
The parameters of the inserted rope are varied until the field lines of the resulting equilibrium match observed features, e.g., threads in a filament or overlying loops. 

A convenient tool for both nonlinear force-free extrapolation and the flux rope insertion method is the so-called magnetofrictional relaxation of the field to force-free numerical equilibria \citep{1981JCoPh..41...68C, 1986ApJ...309..383Y, 1986ApJ...311..451C, 1996ApJ...473.1095R}. 
In a previous paper \citep{2005A&A...433..335V}, we described our implementation of magnetofrictional nonlinear extrapolation and its application to reconstruct a relatively simple force-free magnetic field, which contains an approximately current-neutralized flux rope rooted in the flux concentrations of a simple bipolar magnetogram \citep{2003A&A...406.1043T}. 
That paper included a thorough study of the dependence of the reconstruction quality on the numerical resolution.
In \cite{2007SoPh..245..263V} an improved implementation of the magnetofrictional method allowed us to significantly increase the accuracy in the reconstruction of the well known force-free equilibria by \cite{1990ApJ...352..343L} in comparison to similar extrapolations summarized in \cite{2006SoPh..235..161S}.
Applications of this code to measured magnetograms are included in \cite{2008SoPh..247..269M}, \cite{2008ApJ...675.1637S}, and \cite{2009ApJ...696.1780D}. 

Here we extend the testing of our magnetofrictional code in the reconstruction of solar-relevant solutions of Eqs.~(\ref{eq:fff}) by applying it to the flux rope equilibrium constructed by \citeauthor{1999A&A...351..707T} (\citeyear{1999A&A...351..707T}, hereafter TD).
This configuration describes structures that emerge from the solar interior already twisted and carry a net current \cite[e.g.,][]{1995ApJ...446..877L, 1996ApJ...462..547L, 2000ApJ...532..616W}, and it is in this sense complementary to the one in \cite{2005A&A...433..335V} which represents the case of an originally potential field twisted solely by photospheric displacements. 
The relevance of the TD equilibrium as a model for solar active regions and eruptive configurations was demonstrated by many investigations \citep[\eg][]{2000SPD....31.0236Z,2004A&A...413L..23K,2005ApJ...630L..97T, 2008ApJ...674..586S, 2008A&A...481L..65M}. 
In particular, it has been shown that the TD flux rope can be subject to ideal MHD instabilities \citep{2003ApJ...588L..45R,2004A&A...413L..27T,2007AN....328..743T,2007ApJ...670.1453I}.
Depending on its parameters, the equilibrium can include a \emph{hyperbolic flux tube (HFT)} or \emph{bald patches (BPs)} (see Sect.~\ref{s:tdeq} for detail). 
Such \emph{structural features} are thought to play an important role in the initiation and evolution of coronal mass ejections (CMEs) and their associated flares. 
Moreover, the occurrence of BPs changes the topology of the field. 

Our goal in this paper is to provide a detailed analysis of the reconstruction capabilities of the magnetofrictional code for this test field. 
We address the flux rope morphology, including the twist and the structural features HFT and BP, the energy and helicity contents, and the influence of the initial condition used by the extrapolation code.
We apply the code also to an unstable case, expecting this to be useful in judging extrapolations of magnetograms that were taken during the initial phase of an eruption or immediately prior to its onset. 
Finally, we will compare our results to the previous reconstruction of the TD field by \cite{2006A&A...453..737W}, which employed a different extrapolation scheme. 

Using a force-free test field implies that the magnetogram is compatible with the force-free hypothesis of the extrapolation method.
Therefore, in the following we do not need to consider deviations from force-freeness in the photosphere, the ambiguity of the transverse field direction, noise, and insufficient resolution, all of which are associated with measured magnetograms.
Attempts to deal with some of these problems are discussed, e.g., in \cite{1997SoPh..174..191M}, \cite{1997SoPh..174..129A}, \cite{2006SoPh..233..215W}, \cite{2007A&A...476..349F}, \cite{2008SoPh..247..269M}, \cite{2008SoPh..247..249W}, \cite{2008ApJ...675.1637S}, and \cite{2009ApJ...696.1780D}. 

The paper is organized as follows. 
The TD equilibria used below are described in Sect.~\ref{s:tdeq}, the extrapolation method is briefly outlined in Sect.~\ref{s:method}, and the results are presented and discussed in Sect.~\ref{s:results}. 
Our conclusions are summarized in Sect.~\ref{s:conclusions}. 
%
%
%
\begin{figure}
 \resizebox{\hsize}{!}{\includegraphics{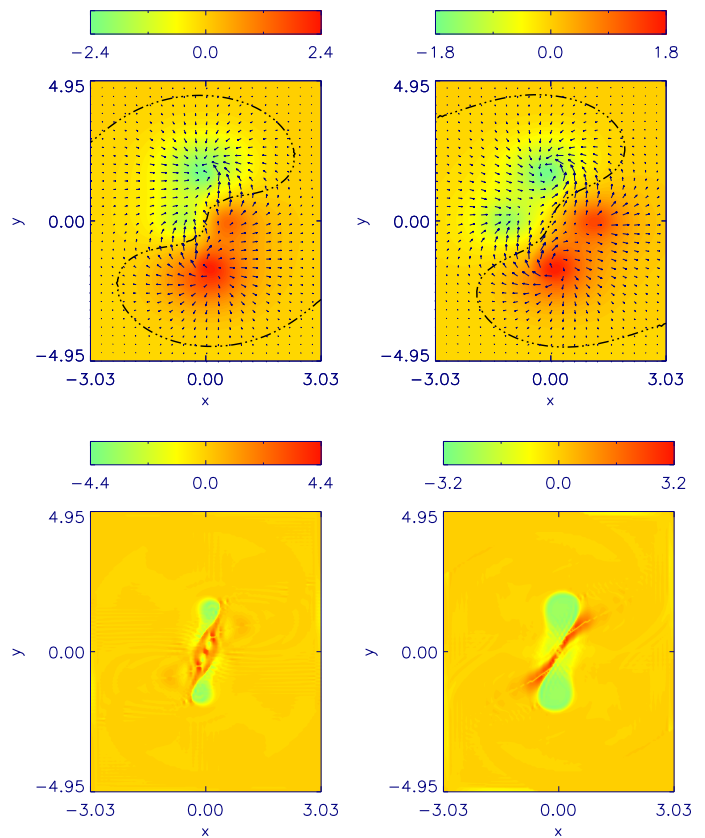}}
 \caption{``Line-of-sight'' magnetogram $B_z(x,y,0)$ with superimposed arrows representing the transverse field \emph{(top)}, and force-free parameter $\alpha(x,y,0)$ \emph{(bottom)}, for the reference fields of the cases High\_HFT \emph{(left)} and BP \emph{(right)}. 
 \label{f:alpha_modb}}
\end{figure}

\section{Test equilibria} \label{s:tdeq}

\citeauthor{1999A&A...351..707T} (\citeyear{1999A&A...351..707T}, TD) constructed a force-free equilibrium of a toroidal current channel to obtain an analytical model of a solar active region that contains free magnetic energy. 
Being force free and carrying a net current, the equilibrium belongs to the class of tokamak equilibria, also known as Shafranov equilibria \citep{1966RvPP....2..103S}, which \emph{require an external poloidal field}, i.e., one due to sources other than the current channel, to balance the always outward directed Lorentz self-force (hoop force) of the bent current channel. 
TD used a pair of magnetic sources, placed symmetrically to the torus plane at the symmetry axis of the torus, to model the external poloidal field, $B_\mathrm{ep}$, of a solar active region. 
They also added a line current running along the symmetry axis to include an external toroidal field, $B_\mathrm{et}$, which permits to model the magnetic shear typically present in active regions and prevents the twist profile from becoming infinite at the surface of the torus. 
The torus is arranged vertically, with its symmetry axis submerged below the plane $\{z=0\}$, which represents the photosphere, resulting in a smooth, divergence-free field in the coronal volume $\{z>0\}$. 
Inside the torus, the field is obtained by modifying the force-free field of a locally straight current channel of prescribed radial profile to match the external potential field smoothly at the surface. 
The original version of the equilibrium uses a uniform radial profile of the current density, resulting, after the matching, in a profile $J(\rho)$ that increases moderately toward the surface at $\rho=a$, where it drops to zero [$J(\rho>a)=0$]. 
The requirement of force-freeness implies that a toroidal field component is always present inside the torus, regardless of the value of $B_\mathrm{et}$. 
The force-freeness within the current channel (i.e., the  equilibrium condition) is satisfied only approximately, but numerical relaxation runs verify that the approximation is very good for high toroidal aspect ratio, $R/a\gg1$ \citep{2004A&A...413L..23K}, and tolerable down to moderate and likely more realistic aspect ratios of order $R/a\sim2$ \citep{2008ApJ...674..586S}. 

If characterized by its field lines, the equilibrium represents an arched, line-tied flux rope that is embedded in a (generally sheared) potential field of arcade topology; the flux rope has a somewhat larger cross-section than the current channel.
The ``line-of-sight'' magnetogram, $B_z(x,y,0)$, possesses four flux concentrations, a pair of ``sunspots'' above the point sources, which are located at $({\pm}L,0,-d)$, and a pair of ``satellite polarities'' in the intersection between the current channel and the ``photosphere'' (Fig.~\ref{f:alpha_modb}). 
(Depending on the parameters chosen for the particular realization of the equilibrium, the satellite polarities can actually encompass more flux than the sunspots.) 
The neutral line of the normal field component in the magnetogram, given by $B_z(x,y,0)=0$, runs approximately parallel to the magnetic flux rope axis beneath its apex and runs relatively straight between the two positive and the two negative flux concentrations, i.e., the ``active region'' is to be characterized as essentially bipolar (for a quadrupolar active region in the classical sense one would require the neutral line to meander substantially between opposite flux concentrations or even to be split, with one or several parts forming a closed loop). 

\begin{figure}
 \resizebox{.9\hsize}{!}{\includegraphics{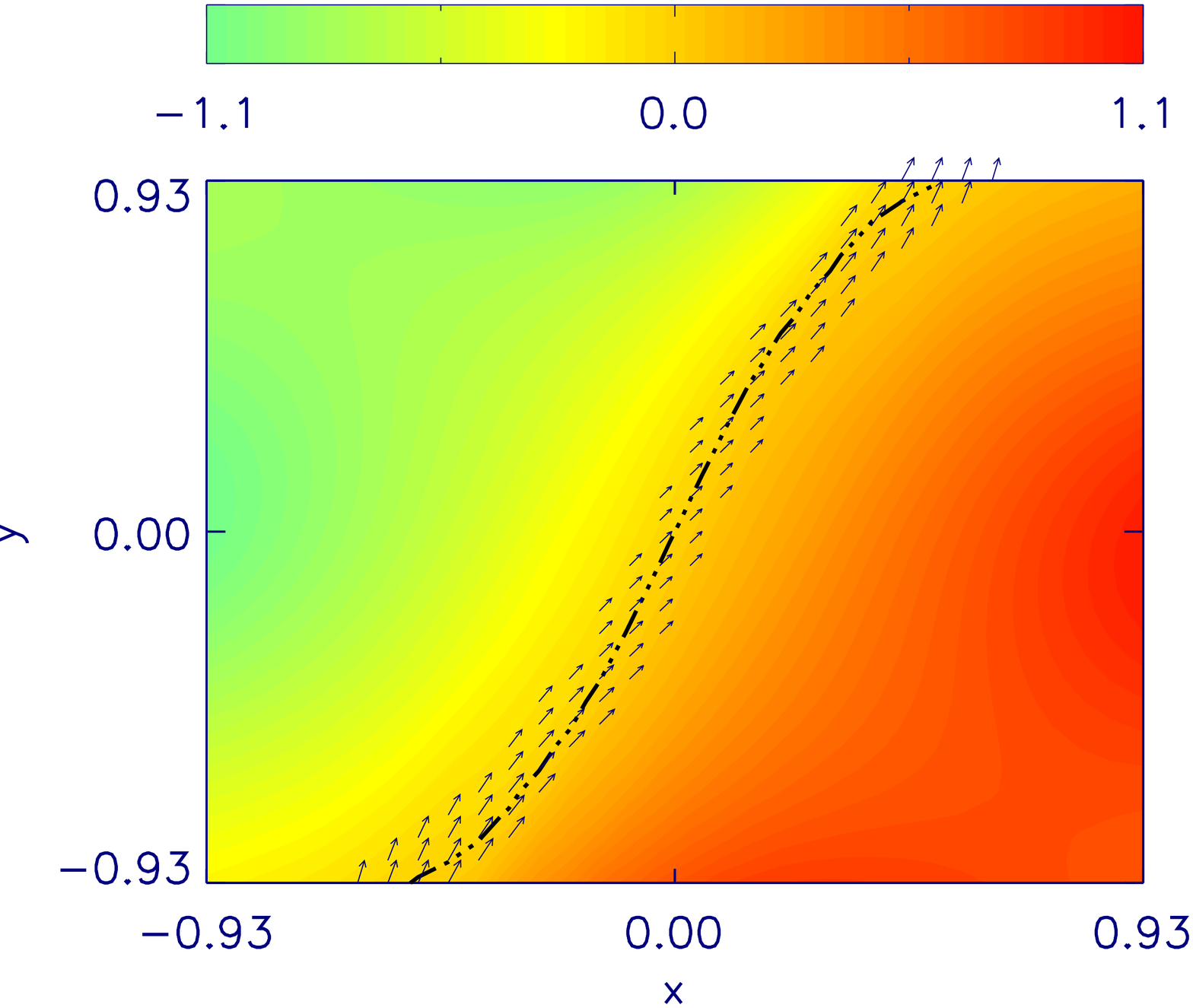}}\\[6pt]
 \resizebox{.9\hsize}{!}{\includegraphics{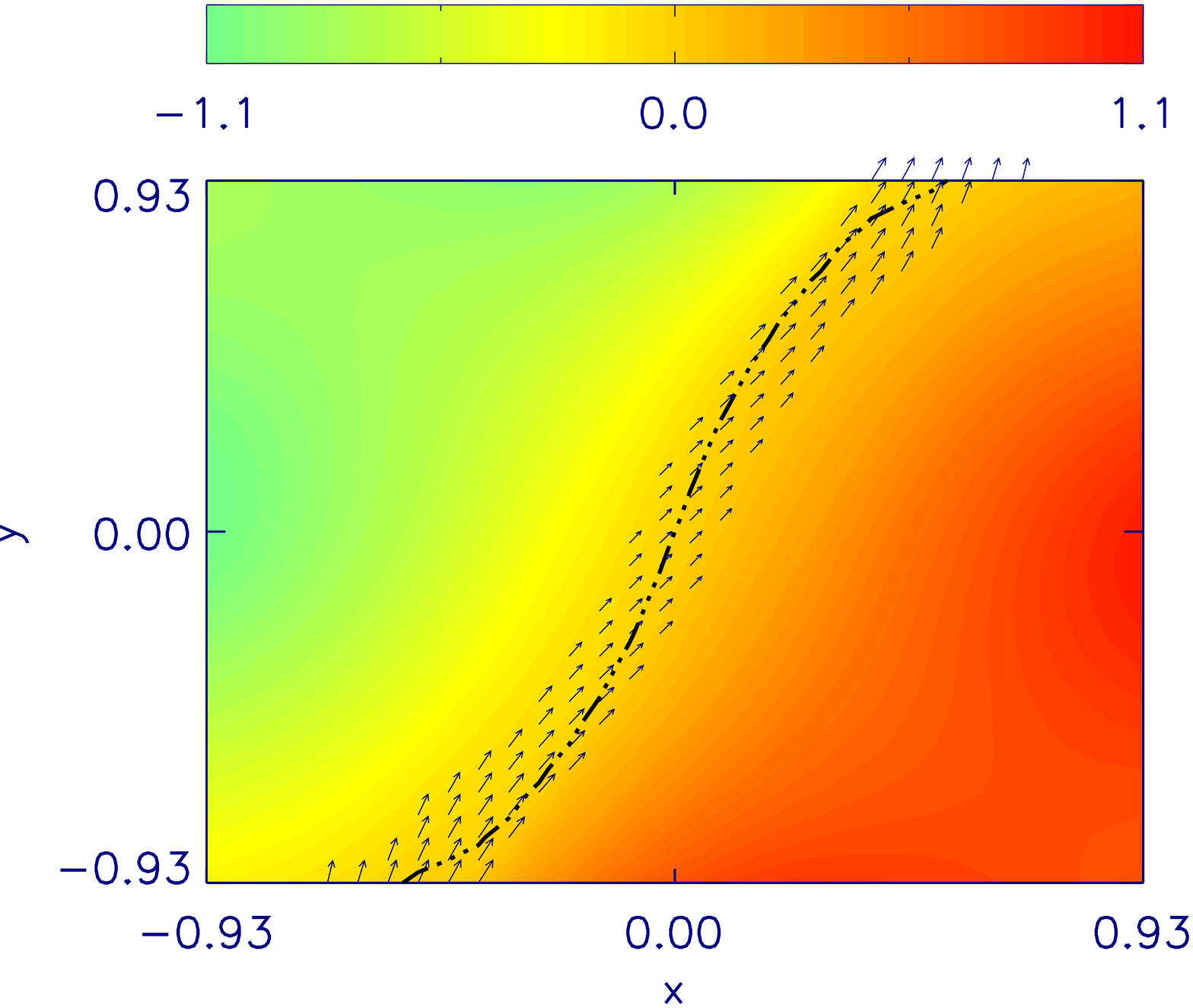}}
 \caption{Contours of $B_z(x,y,z=\Delta)$, with superimposed arrows representing the transverse field in the vicinity of the neutral line of $B_z$, for the reference \emph{(top)} and reconstructed \emph{(bottom)} BP cases. Note the inverse crossing of the neutral line, the characteristic signature of a BP, in the middle of the plots. 
 \label{f:isobp}}
\end{figure}
In a wide range of parameter space, approximately bounded by the conditions that $d<R-a$ and $B_\mathrm{et}<B_\mathrm{ep}$ under the rope, the interface between the flux rope and the surrounding arcade of field lines is a separatrix surface (or quasi-separatrix layer), across which the field line mapping varies abruptly (or rapidly) \citep{1995JGR...10023443P, 1996A&A...308..643D}. 
In the following we will collectively refer to these as \emph{separatrix}. 
They introduce one or both of the following structural features, depending upon the parameters of the equilibrium. 
If the separatrix intersects itself under the flux rope, a line of X-type topology results, which is a concentric ring of radius smaller than $(R-a)$ lying in the torus plane $\{x=0\}$. 
The flux bundle in the vicinity of this line is often referred to as hyperbolic flux tube, HFT \citep{1999ESASP.448..715T, 2002JGRA..107.1164T, Titov2007}. 
The HFT pinches under external perturbations \citep{2003ApJ...582.1172T, 2003ApJ...595..506G}; thus it plays the role of a seed for the vertical flare current sheet in eruptions \citep{2004A&A...413L..27T} and is important for the CME-flare relationship. 
If the separatrix touches the photosphere tangentially, it does so at the neutral line and at these locations the field points in the ``inverse'' direction (from the negative-polarity side of the neutral line to the positive-polarity side, see Fig.~\ref{f:isobp}). 
Such features are referred to as bald patch, BP \citep{1986SoPh..105..223S, 1993A&A...276..564T, 1996A&A...308..233B, 1999A&A...351..707T}.
They have been suggested to be sites where partial eruptions originate and where sigmoidal coronal sources form \citep{2004ApJ...617..600G, 2009ApJ...700L..83G}.

The bipolar structure of the external field implies the absence of null points. 
These relevant topological features \cite[e.g.,][]{1998ApJ...502L.181A} can be included by generalizing the external field, however, we leave this step for future work.

With regard to the reconstruction task, it is important to note that TD equilibria which contain a separatrix are structurally quite different from any potential field that can be constructed from their corresponding line-of-sight magnetogram. 
The potential field will not include a flux rope. 
In the majority of cases, HFT or BP(s) will not be present either, and in case they happen to exist, they will have very different locations than in the TD field. 
At the same time, these equilibria appear to be the simplest force-free configurations currently considered in coronal physics which are structurally (and in the presence of BP(s) even topologically) different from a sheared-arcade field. 
They consist of a single current channel and the minimum two flux systems required: the flux in the rope and the external poloidal flux. 
These are the generic elements of more complex force-free configurations that carry a net current. 
The requirement of a structural or even topological change makes the reconstruction qualitatively different from reconstructions of Low \& Lou fields or of fields obtained by shearing or twisting the photospheric flux with smooth profiles of substantial symmetry \cite[\eg][]{2003A&A...406.1043T}. 
Therefore, the reconstruction of the TD field is a major and necessary step in verifying the capabilities of nonlinear extrapolation schemes. 

\begin{figure*}
 \vspace*{0.5cm}
 \flushleft \ \hspace{10mm} {\bf (a) \hspace{80mm}  \bf (b)} \\ \centering
 \resizebox{0.45\hsize}{!}{\includegraphics[clip=true]{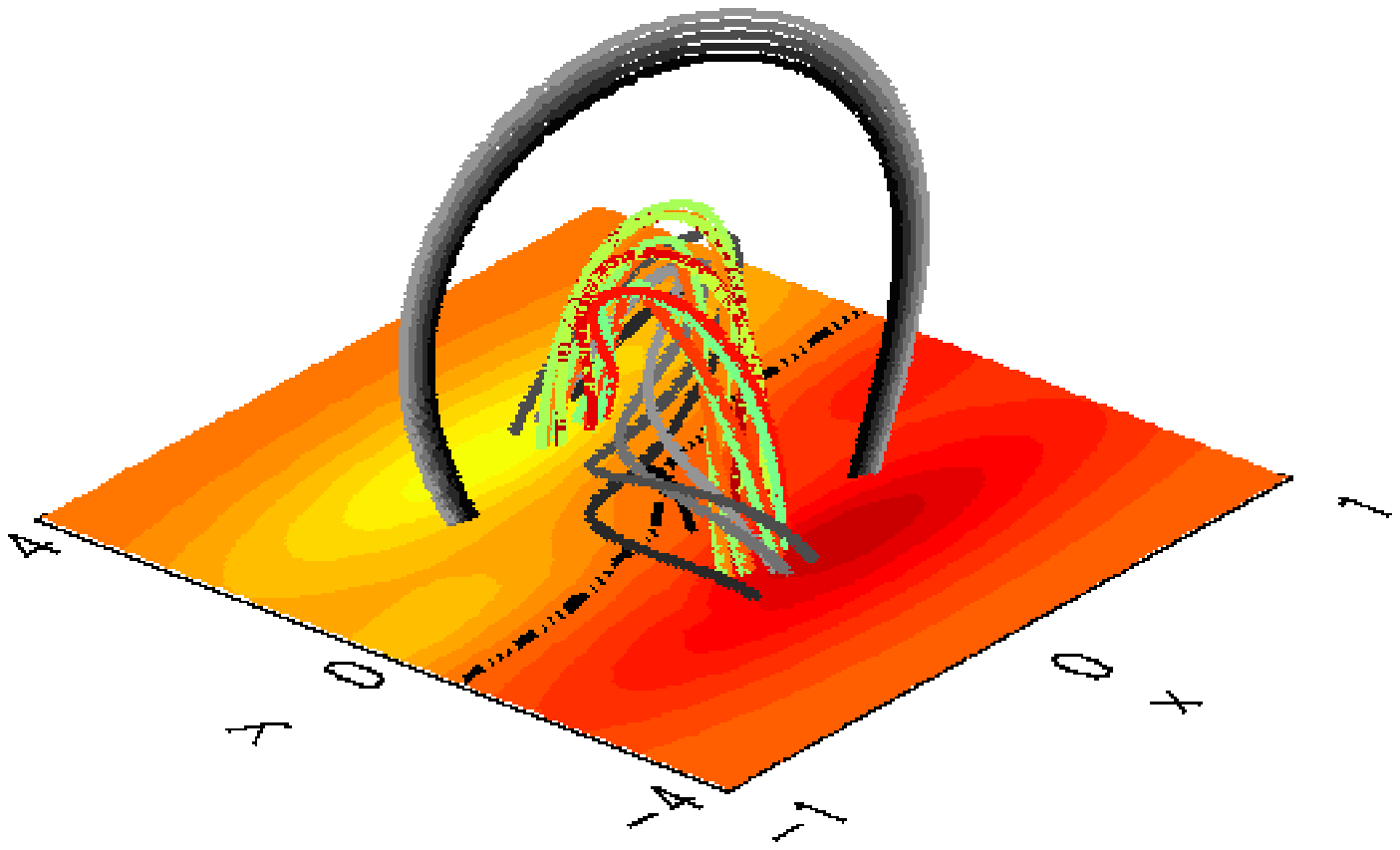}}
 \resizebox{0.45\hsize}{!}{\includegraphics[clip=true]{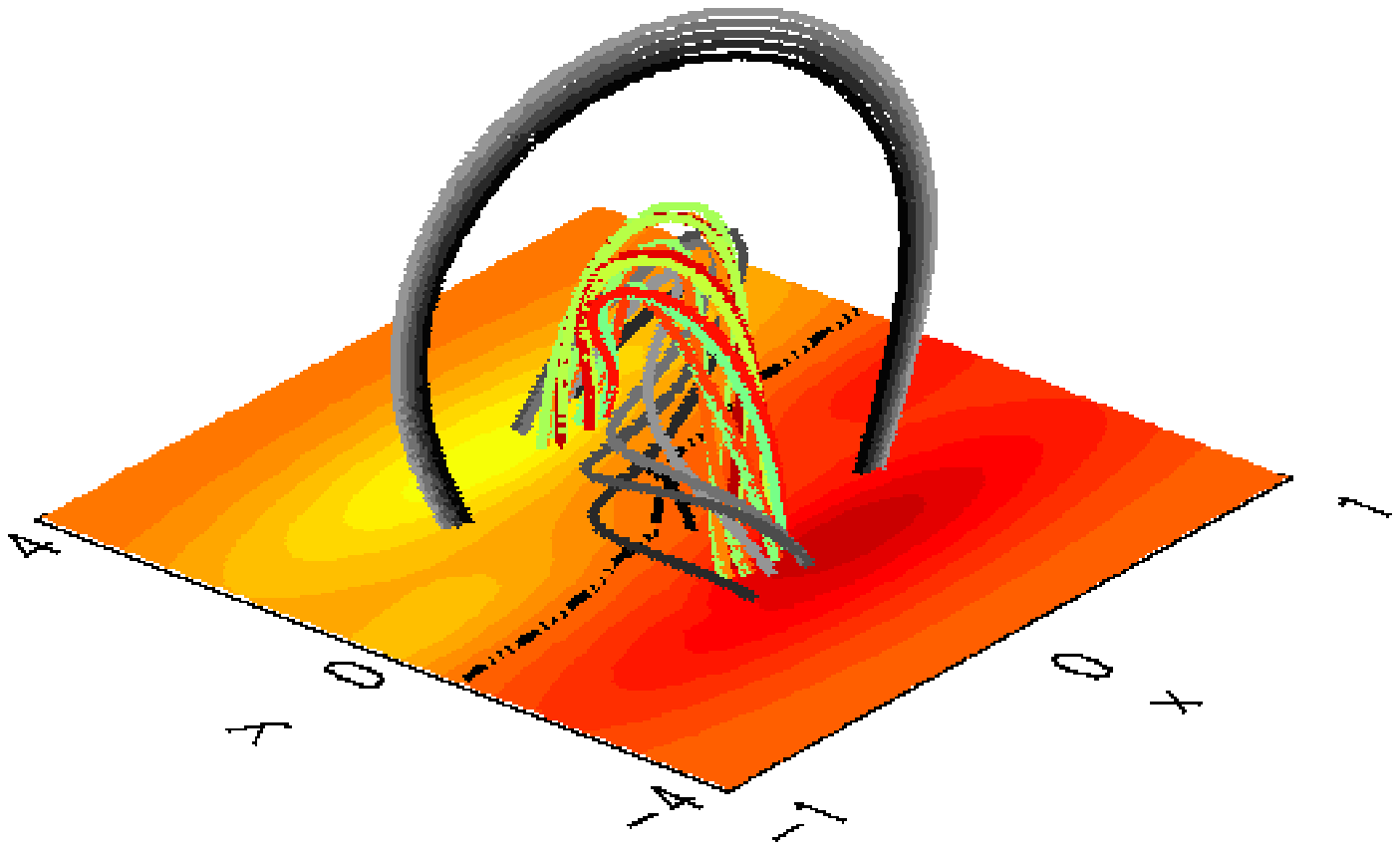}}  \\
 \flushleft \ \hspace{10mm} {\bf (c) \hspace{80mm}  \bf (d)} \\ \centering
 \resizebox{0.45\hsize}{!}{\includegraphics[clip=true]{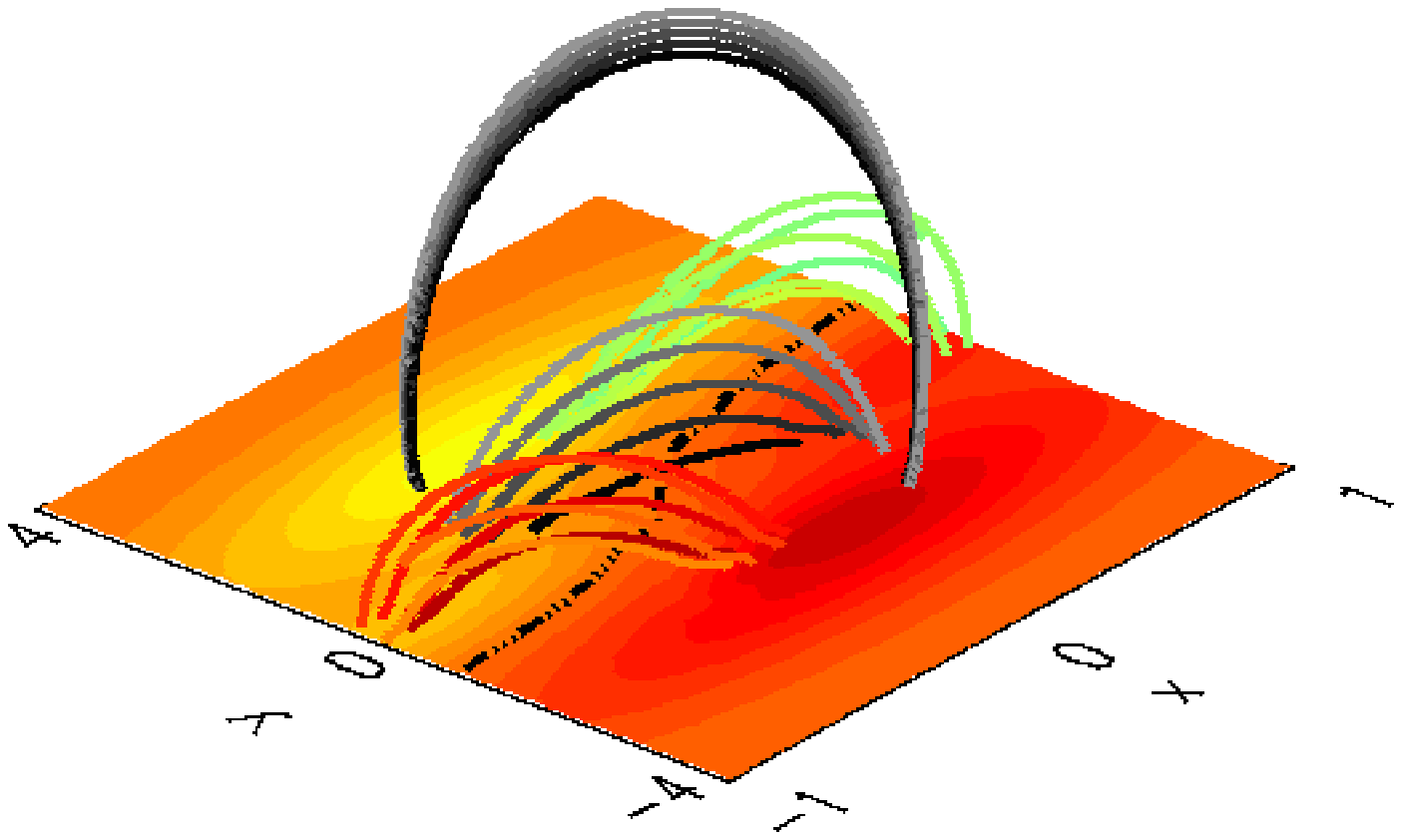}}  
 \resizebox{0.45\hsize}{!}{\includegraphics[clip=true]{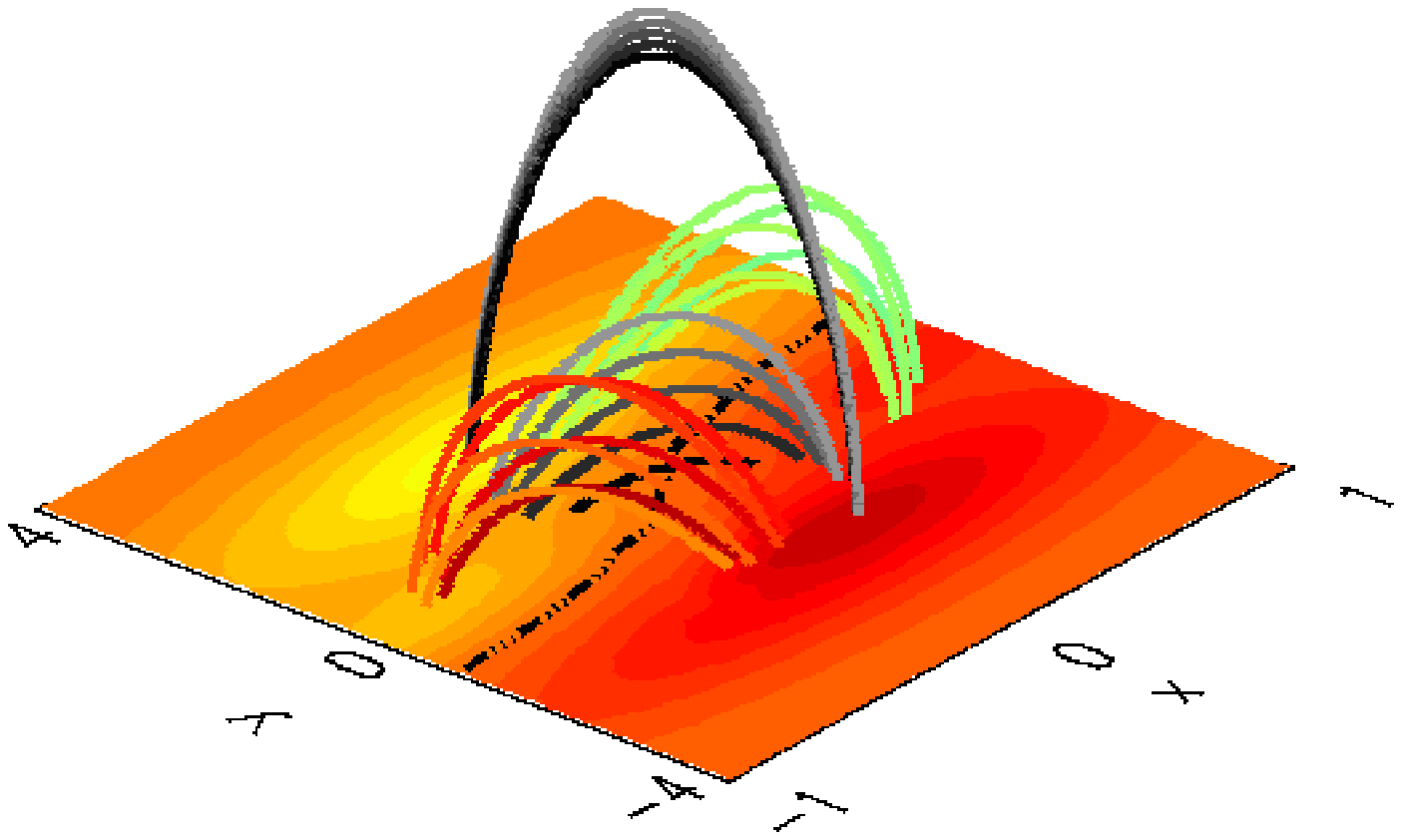}}
 \caption{Selected field lines for the Low\_HFT case; {\bf (a)} reference, {\bf (b)} extrapolated, {\bf (c)} potential, and {\bf (d)} linear field (with $\alpha=-0.2$). 
In {\bf (a)} and {\bf (b)} the flux rope is depicted by two groups of field lines (in red and green), each one  starting from six uniformly spaced points on a circle of radius 0.1, centered at the intersections of the magnetic axis of the flux rope (CFL) with the $z=0$ plane.
Grey field lines are started from the $z$-axis and visualize the potential field above and below the current channel.
In {\bf (c)} and {\bf (d)} the field lines are started from the same set of footpoints as in {\bf (a)} and {\bf (b)}.
Contour plots of $B_z(x,y,0)$ and its neutral line $B_z(x,y,0)=0$ (dot-dashes) are shown in the bottom plane of each panel.
 \label{f:fl_ex14}}
\end{figure*}
An impression of the difficulty faced by the extrapolation code can be obtained from the field line picture, Fig.~\ref{f:fl_ex14}. 
If an HFT is present, the flux connecting the satellite polarities must ``penetrate'' the flux connecting the sunspots in the course of the reconstruction. 
This is due to the fact that, in the potential field that is used as initial condition for the extrapolation, the flux in the satellite polarities connects to the corresponding sunspot of opposite polarity, resulting in two main flux bundles lying side by side. 
In the final TD field, however, the connection between the four flux concentrations in the magnetogram is like a cross (which can be traced back to the requirement that a non-neutralized current channel in force-free equilibrium must be stabilized by an external poloidal field). 
This results in magnetic connections between the sunspots below and above the flux rope, i.e., the flux rope 
passes in between these two parts of the sunspot flux (this can be seen in Fig.~\ref{f:fl_ex14} and, more clearly, in Fig.~\ref{f:fl_erup_nl} below). 
If a BP is present, it is obvious that the field lines in and immediately above it must reverse the direction of neutral line crossing in proceeding from the initial potential field to the solution. 
This implies that field lines of the initial potential field passing low over the BP location, which are simple short loops rooted in the vicinity of the BP, must be replaced by field lines that are typically long and rooted at very different locations. 

We note that the configuration considered in \cite{2008SoPh..247..269M} represents an intermediate case between purely sheared configurations (like, e.g., the family of Low \& Lou fields) and the TD field. 
It was obtained by means of the flux rope insertion method such that the magnetic axis of the rope passes at a low height of only a few Mm above the photosphere along much of the length of the rope \citep{2004ApJ...612..519V}. 
This corresponds to a strongly submerged TD rope with $R-a<d<R+a$ and contains a separatrix which may or may not form an HFT or a BP. 
The configuration used in \cite{2008SoPh..247..269M} includes a BP, but the extrapolations presented in that paper were not checked for the reconstruction of this feature.

We have searched published extrapolations of solar magnetogram data for the occurrence of BPs (inverse field direction at the neutral line) and of the flux penetration feature characteristic of an HFT. 
There appear to be only two cases of a BP \citep{2009ApJ...693L..27C, Guo&al2010} and one case of flux penetration \citep{2004A&A...425..345R}. 

For the purposes of the present paper, the original TD equilibrium is modified in two ways. 
First, in place of the line current, we use a pair of dipoles as sources of the external toroidal field \citep{2005ApJ...630L..97T}. 
This yields a compact distribution of magnetic flux in the magnetogram plane and a steeper decrease of the external field with height than the original TD field, both of which are more realistic with regard to solar active regions. 
The dipoles are placed below the footpoints of the flux rope and point nearly vertically, such that the field lines connecting them through the torus section in $\{z>0\}$ match the geometry of the torus section closely. 
The conservation of toroidal flux in the rope now requires the minor radius of the torus to vary along its length. 
Since $B_\mathrm{et}$ still points approximately in the direction of the current, its strength remains a free parameter to a good approximation for parameters of interest here, similar to the original equilibrium. 
The break of the toroidal symmetry also causes the magnetic axis of the rope and the HFT to bulge out of the plane $\{x=0\}$, but the effect remains small, since $B_\mathrm{et}$ varies only moderately along the coronal flux rope for the chosen considerable depth of the dipoles. 
Second, the radial profile of the current density is changed such that it no longer peaks at the surface but in the middle of the channel, and decreases monotonically towards zero at the surface. 
This removes unrealistically steep gradients of the current density at the surface of the torus, including the magnetogram plane. 
The expressions for the equilibrium with this parabolic current density profile will be provided in a future publication (V.~S.~Titov, in preparation). 

\begin{table}
\caption{Parameters of the TD equilibria}
\label{t:tdparam}
\centering
\begin{tabular}{c c c c c c c}
\hline \hline
Case         & $R$  & $a$  & $d$  &  $L$ & $z_0$  & $m$    \\
\hline
{High\_HFT}  & 1.83 & 0.67 & 0.83 & 0.50 & -1.75  & -65.82 \\ 
{Low\_HFT}   & 1.83 & 0.67 & 0.83 & 0.83 & -1.75  & -65.96 \\ 
{No\_HFT}    & 1.83 & 0.67 & 0.83 & 1.17 & -1.75  & -65.80 \\ 
{BP}         & 1.83 & 0.90 & 0.83 & 1.17 & -1.75  & -38.51 \\ 
{Unstable}   & 1.83 & 0.41 & 0.83 & 0.83 &  5.00  & -186.7 \\ 
\hline
\end{tabular}
\flushleft
\footnotesize{Normalized parameters defining the TD equilibria used in the paper: $R$ and $a$ are the torus major and minor radii, respectively, $d$ is the depth of the torus center, $L$ is the distance of the magnetic charges to the center of the torus, and $z_0$ and $m$ are the depth and strength of the magnetic dipoles, respectively.}
\end{table}
We consider the modified TD equilibrium for several parameter sets, four stable configurations and an unstable one.
The parameters defining each equilibrium are given in Table~\ref{t:tdparam}. 
All variables are normalized by a characteristic length, taken here to be the apex height of the geometrical torus axis, $R-d$, by the magnetic field strength at the apex point, and by quantities derived thereof.

Three of the four stable equilibria have identical geometric flux rope parameters and an almost identical left-handed, average end-to-end twist of $\approx 2.1 \pi$ in the center of the current channel (see Sect.~\ref{s:results} for the averaging procedure), but different properties of the ambient field. 
In the first two cases an HFT is present, staying close to the photosphere in one (Low\_HFT) and reaching significantly into the volume in the other (High\_HFT). 
A third case (No\_HFT) has no HFT above the photospheric plane, hence, it has a simpler magnetic structure than the cases with an HFT. 
Due to the relatively strong toroidal field component chosen, these three configurations do not have bald patches.
The fourth stable case (BP) has a left-handed average twist of $\approx1.8\pi$, close to the twist of the first three equilibria, but the minor radius of the torus is enlarged (and $B_\mathrm{et}$ correspondingly reduced) to introduce a bald patch in the resulting field. 

Figure~\ref{f:alpha_modb} illustrates for the High\_HFT and BP cases that the field strength in the magnetogram decreases rapidly with distance from the four flux concentrations. 
The distribution of $\alpha(x,y,z=0)$ in the same figure shows that the field is highly nonlinear (see below for the occurrence of currents outside the torus in these plots). 
The transverse field near the section of the neutral line between the flux concentrations is strongly aligned with the neutral line for the BP configuration, as in a filament channel on the Sun. 
A zoom into this area is plotted in Figure~\ref{f:isobp}, demonstrating by the inverse crossing of the neutral line in the central part of the magnetogram that a BP is indeed present. 

To obtain an unstable case, we increase the average twist to $\approx 2.7 \pi$ by reducing the minor torus radius. 
This configuration is unstable with respect to the helical kink instability. 

The analytical expressions in TD and their modifications described above represent only approximate equilibria. 
Moreover, the numerical determination of the current density introduces additional (``discretization'') errors. 
Therefore, the individual magnetic configurations were first relaxed to numerical equilibria, using the MHD code described in \cite{2003A&A...406.1043T} on uniform Cartesian grids with a resolution $\Delta=0.06$ in all directions. 
This introduces only little change in the shape of the flux rope, but new layers of current density, not present in the analytical model, form in the separatrix and its vicinity, primarily below the current channel. 
These are visible in the $\alpha(x,y,0)$ plots of Fig.~\ref{f:alpha_modb} as narrow strips of forward and reverse current between the footpoints of the channel. 

From each numerically relaxed stable TD equilibrium, a data cube of size $[-3.03, 3.03] \times [-4.95, 4.95] \times [-0.06, 4.44]$ was extracted, containing the nonlinear magnetic field to be used as the test field for the extrapolation code (the somewhat larger cube used for the unstable case is given in Sect.~\ref{s:unstable}). 
We refer to these data cubes as the \emph{reference field}, $\vB^\mathrm{ref}$, for the given set of parameters. 
The reference fields were then reconstructed by magnetofrictional extrapolation, employing the same grid as in the MHD relaxation.
The High\_HFT reference field was also constructed and reconstructed using a uniform grid resolution of 0.03 in each direction. Since the achieved accuracy of reconstruction turned out to be comparable to the one obtained with resolution of 0.06, we selected the latter value for the present study.
Although the nonlinear solution is known in the whole discretized volume, and specifically on its lateral and top boundaries, the vector field in the bottom plane, $\vec{B}^\mathrm{ref}(x,y,z=0)$, is the only information used in the reconstruction of all test fields, because this is the challenge that the extrapolation codes have to face when they are confronted with observed magnetograms. 

\begin{table}
\caption{Compatibility of magnetogram and extrapolation method
\label{t:mgmff}}
\centering
\begin{tabular}{c c c c}
\hline \hline
{Case}  & {$\epsilon_{flux} \times 10^{8}$ }  &   { $\epsilon_{force} \times 10^{5}$  }   &  {$\epsilon_{torque} \times 10^{5}$}\\ 
\hline 
High\_HFT& -0.99 &  2.60 &  0.93 \\
Low\_HFT & -1.83 &  5.84 &  1.24 \\
No\_HFT  &  2.28 &  8.56 &  1.66 \\
BP       &  3.52 &  29.71&  2.47 \\
Unstable & -4.21 &  90.42&  0.12 \\
\hline 
\end{tabular}
\flushleft
\footnotesize{Normalized flux, force, and torque imbalance \cite[as defined in][]{2006SoPh..233..215W} of the five magnetograms analyzed.  }
\end{table}
Since the reference TD fields satisfy Eqs.~(\ref{eq:fff}) only to the accuracy of the finite-difference scheme used in the relaxation, it is necessary to quantify the compatibility of their magnetograms with the force-free hypothesis implicit in the extrapolation method. 
For this purpose, we use normalized integral averages of flux, force, and torque in the lower boundary as defined in \cite{2006SoPh..233..215W}, which vanish for perfect compatibility. 
These metrics, reported in Table~\ref{t:mgmff}, show that the flux imbalance is negligible and that the residual force and torque values remain small, even for the unstable case. 

%
%
\section{Extrapolation method}\label{s:method} 

The implementation of magnetofrictional extrapolation used in this paper was described in detail in \cite{2005A&A...433..335V, 2007SoPh..245..263V}, so that we can restrict ourselves here to a short outline of the method and its recent improvements. 
Our formulation of the method consists in a pseudo-temporal evolution of an initial field subject to the equation
\begin{eqnarray}
\frac{\partial \vB}{\partial t}=\Nabla \times (\vv \times \vB) +c_L \Nabla (\divb), \label{eq:mfeqs} \\
\mbox{where} \qquad \vv := c_Y\frac{\vJ \times \vB}{B^2}, \qquad \vJ = \Nabla \times \vB \label{eq:mfdef},
\end{eqnarray}
and $c_Y$ and $c_L$ are numerical factors fixed by suitable stability criteria.
The initial field is usually chosen to be a potential field derived from the ``line-of sight'' magnetogram $B_z^\mathrm{ref}(x,y,0)$, with the full vector magnetogram $\vB^\mathrm{ref}(x,y,0)$ overwritten in the photospheric layer.
The rightmost term in  Eq.~(\ref{eq:mfeqs}) is the $\divb$ cleaner (see, \eg \citeauthor{Keppensetal2003} [\citeyear{Keppensetal2003}] and references therein).
By applying the divergence operator, it is readily seen that Eq.~(\ref{eq:mfeqs}) reduces to a diffusion equation for the scalar field $\divb$. 
Since also the first term on the right hand side of Eq.~(\ref{eq:mfeqs}) is of parabolic type \citep{1986ApJ...311..451C}, Lorentz forces and errors of $\divb$ diffuse out of the computation domain through the boundaries in the course of the extrapolation.
This is best seen by computing the time derivative of the magnetic energy $E_\mathrm{mag}=\int dV\,B^2/2$.
Multiplying Eq.~(\ref{eq:mfeqs}) with $\vB$ and integrating by parts, it is found that 
\begin{equation}
\partial_t E_\mathrm{mag} =-c_Y\int dV \ |\vJ_\perp|^2 -c_L \int dV \ (\divb)^2, 
\label{eq:mfintegral}
\end{equation}
which implies that the magnetic energy decreases as long as there are Lorentz forces and magnetic sources in the domain.
Here it has been assumed that the boundaries are perfectly conducting \cite[as in][]{1986ApJ...311..451C} and that the field is solenoidal at them, so that surface integrals do not contribute. 
During the pseudo-temporal evolution, the decrease of the perpendicular current density $\vJ_\perp$ reduces also the relaxation velocity $\vv$ [Eq.~(\ref{eq:mfdef})].
If a static state is reached, then the field is divergence-free and $\vv=0$ implies that it is also force-free.
By construction, such a static solution extends smoothly from the vector magnetogram in the bottom boundary and represents the extrapolated nonlinear force-free field.
We note that by overwriting the vector magnetogram in $\{z=0\}$, we actually introduce a finite jump in the field close to the photosphere, yielding  Lorentz forces and errors in the solenoidal property; consequently, surface integral terms in Eq.~(\ref{eq:mfintegral}) are important initially. 
This results in an initial rise of the energy not described by Eq.~(\ref{eq:mfintegral}). 
A discussion of this effect will be given in a future paper.

Equation~(\ref{eq:mfeqs}) can be formally derived from the MHD equations \citep[as in][]{1986ApJ...309..383Y} by assuming that viscosity dominates the evolution (hence the characterization of the method as ``magnetofrictional'').
However, since such a strong viscosity has no immediate physical justification for the nearly collisionless 
coronal plasma, one can equivalently consider Eq.~(\ref{eq:mfeqs}) to be a numerical tool that finds a force-free magnetic field for given boundary conditions, with only the initial potential field (before the vector magnetogram is overwritten) and the final 
nonlinear force-free field having a well defined physical meaning. 

Our implementation of the magnetofrictional code employs a fourth-order central-difference scheme for the space discretization.
The pseudo-time discretization is based on a general MHD-stability prescription \citep[see details in][]{2007SoPh..245..263V}.
In order to reduce the execution time, the time discretization was recently complemented by a Runge-Kutta-Chebyshev acceleration technique \citep{1994alexiades,2000IMA.120...19E}.
The detailed description and testing of this upgraded pseudo-time discretization is beyond the scope of the present paper and will be provided in a future publication.

The field in the photospheric boundary is fixed to keep the input magnetogram values throughout the whole extrapolation.
On the side and top boundaries of the extrapolation volume, which is identical to the volume of the reference field given in Section~\ref{s:tdeq}, the normal component is derived from the inner field values so to have vanishing $\divb$ in the boundary. 
The transverse component is given by a fourth-order polynomial extrapolation of the inner field values onto the boundaries.
These boundary conditions were tested in \cite{2007SoPh..245..263V}, where more detail can be found. 
Finally, all extrapolations presented here start from the potential field obtained by Schmidt's method \citep{1964NASSP..50..107S, 1989SSRv...51...11S}, except for a potential-field and a linear-field case discussed in Sect.~\ref{s:inicond}, for which the method by \cite{1978SoPh...58..215S} was used. 

\section{Analysis of the extrapolation results}\label{s:results}

The reconstructed fields, $\vB(x,y,z)$, are compared to their respective reference field, $\vB^\mathrm{ref}(x,y,z)$, in an ``analysis domain'' given by clipping 
20 grid layers from the top and lateral boundaries of the extrapolation volume. 
This is done in recognition of the fact that the magnetograms have considerable margins of current-free field around the flux rope footpoints. 
The currents remain relatively small in the corresponding parts of the extrapolation volume, where the relaxation is slowed and, therefore, the quality of reconstruction is reduced. 
Similar practice, with larger margins, was adopted in \citet{2006SoPh..235..161S} and \citet{2008SoPh..247..269M}. 
The quality of the reconstructed equilibria is quantified in the following in several ways.

The degree of force-freeness is measured by the current-weighted, average sine-angle between current and magnetic field, 
\begin{equation}
\sj=\frac{\sum_iJ_i \sigma_i}{\sum_i J_i }, \qquad \mbox{where } \sigma_i = \frac{\vB_i \times \vJ_i}{B_iJ_i} 
\label{eq:sj}
\end{equation}
and the sums run over all grid nodes $i$ in the analysis domain.  
This same quantity is sometimes referred to as $CWsin$ \citep[\eg][]{2008ApJ...675.1637S}.
The solenoidal property is quantified, following \cite{2000ApJ...540.1150W}, by the average over the grid nodes $\langle\,|f_i|\,\rangle$ of the fractional flux 
\begin{equation}
f_i = \frac{\int_\mathcal{V} d\mathcal{V} \ \divb_i}{\int_{\partial\mathcal{V}} \vec{dS} \cdot \vB_i},  
\label{eq:fi}
\end{equation}
through the surface $\partial\mathcal{V}$ of a small volume $\mathcal{V}$ including the node $i$. 
For uniform grid spacing $\Delta$, this reduces to $f_i= \Delta (\divb_i) /(6 |\vB_i|)$ and $\langle\,|f_i|\,\rangle=\sum_i |f_i|/N$, where $N$ is the number of grid points. 
The influence of the residual spurious magnetic charges implied by $\divb\ne0$ can also be quantified by the non-solenoidal part of the magnetic field, $\vB_\mathrm{ns}$. 
Using Gauss' theorem, we have 
\begin{equation}
\vB_\mathrm{ns}= \frac{\Delta^3}{4\pi}\sum_{i} (\nabla \cdot \vB)_i\frac{\vx-\vx_i}{|\vx-\vx_i|^3}\,,
\label{eq:bns}
\end{equation}
which is evaluated on an auxiliary grid displaced by $\Delta/2$ in each direction. 
All three quantities should be as small as possible. 
We have demonstrated in \cite{2007SoPh..245..263V} that our code reduces the value of $\divb$ to machine precision in some cases; however, due to the parabolic nature of Eq.~(\ref{eq:mfeqs}), this may require long execution times.
For the purposes of the present work, we terminated the relaxation when the force-freeness metric $\sj$ tended to reach a plateau of sufficiently small value (of order $10^{-2}$) and $\langle\,|f_i|\,\rangle$ was of the same order of magnitude as the corresponding value for the initial potential field or smaller. 

Since the reference field $\vB^\mathrm{ref}$ is known, we can perform a direct comparison between $\vB^\mathrm{ref}$ and the reconstructed field $\vB$ obtained from the magnetogram.
Several comparison metrics were defined in \cite{2006SoPh..235..161S}, out of which we adopt the mean vector error
\begin{equation}
E_\mathrm{M} = \frac{1}{N}\sum_i\frac{|\vB^\mathrm{ref}_i-\vB_i|}{|\vB^{ref}_i|}\,.
\label{eq:em}
\end{equation}
The smaller $E_\mathrm{M}$, the better the reconstruction. 
As for previous reconstructions of test fields \citep{2006SoPh..235..161S, 2007SoPh..245..263V, 2008SoPh..247..269M}, this metric is the most sensitive one to reconstruction errors of the TD field. 
The further metrics of direct comparison defined in \citet{2006SoPh..235..161S}, $C_\mathrm{vec}$, $C_\mathrm{CS}$, $E_\mathrm{N}$, yield even smaller deviations from the ideal value (1 or 0) than $E_\mathrm{M}$ for all cases analyzed here. 

To judge geometry and topology, we first determine the apex height of the flux rope's magnetic axis, also referred to as the ``central field line'', CFL, in the following. 
(The CFL differs slightly from the geometrical axis of the torus, since the external potential field drops with height and, hence, contributes differently to the bottom and top parts of the flux rope.) 
Next we check for the presence of an HFT and its apex height. 
Both apex heights are given by inversion points of the poloidal component of the field at the line-symmetric $z$ axis. 
This is well approximated by $B_x(0,0,z)$ because the flux rope writhes only rather weakly out of the plane $\{x=0\}$ in the equilibria considered in this paper. 
Finally, we check whether the BP in the fourth stable test field is recovered at the correct section of the neutral line. 

The TD equilibrium can be susceptible to the helical kink instability \citep{2004A&A...413L..27T} and to the torus instability \citep{2006PhRvL..96y5002K}. 
The parameters controlling these instabilities should be reliably and accurately reproduced by the extrapolation.
The main parameter which regulates the stability with respect to the helical kink mode is the field line twist.
This quantity, $\Phi = lB_\Phi(\rho)/(\rho B_\zeta(\rho))$, is defined with respect to the CFL of length $l$ using local cylindrical coordinates $(\rho,\phi,\zeta)$. 
We use its average in a circular cross section of normalized radius 0.3 centered at the apex of the CFL. 
Within this radius, the average twist is $\langle\,\Phi\,\rangle \approx 2.1\,\pi$ for three of the four stable equilibria (High\_HFT, Low\_HFT, and No\_HFT) and $\approx 1.8\, \pi$ for the fourth stable case (BP). 
For the unstable case discussed in Sect.~\ref{s:unstable}, the average twist in $\rho\le0.3$ is $\approx 2.7\,\pi$.
The threshold of the torus instability is given in terms of the ``decay index'' of the external poloidal field, 
$n = - R\,d \ln B_\mathrm{ep}/dR$, where $R$ is the major torus radius. 
For $n>n_\mathrm{cr}$, with the threshold lying in the range $n_\mathrm{cr}\approx3/2\dots2$ \citep{2006PhRvL..96y5002K}, an expansion of the current ring out of equilibrium 
cannot be balanced by the external poloidal field. 
Computing the index $n$ requires the knowledge of the \emph{external} poloidal field, which is not immediately available for our numerical solutions, since it cannot be easily separated from the poloidal field produced by the ring current.
Therefore, we compute the decay index of the total poloidal field at the $z$ axis, given to a good approximation by $B_x(0,0,z)$, 
\begin{equation}
\tilde n = \left\langle-z\frac{d \ln B_x(0,0,z)}{dz}\right\rangle.
\label{eq:tilden}
\end{equation}
The average is taken in a small interval in $z$, which is located just above the current channel in all equilibria considered here.
The quantity $\tilde n$ is not suitable to determine the onset of the torus instability, 
but it serves the purpose of quantifying how well the stabilizing potential field above the current channel is reproduced by the extrapolation. 

The magnetic energy, $E_\mathrm{mag}$, and relative magnetic helicity, $H_\mathrm{mag}$, should also be accurately reproduced by the reconstructed field.
We use the formulae by \cite{2000ApJ...539..944D} to obtain an approximation of  $H_\mathrm{mag}$ for both the reference and the reconstructed fields.

In order to reliably compare the effects that topological differences have on the extrapolation quality, all computations presented in this paper for stable TD equilibria were performed using identical numerical parameters, rather than being individually optimized for best reconstruction. 

%
%
%

\begin{table}
\caption{Reconstruction accuracy of the stable TD equilibria}
\label{tb:potvgerr}
\centering
\begin{tabular}{ccccc}
\hline \hline
Figure of merit  & {High\_HFT}  &   {Low\_HFT}   &  {No\_HFT} &{BP}\\
\hline
$\sigma_J \times 10^{2} $ & 1.55     &  0.73    &   1.18    &  0.80   \\
$\langle\,|f_i|\,\rangle \times 10^{5}$ & 6.04     &  6.90    &  10.28    & 10.05   \\
$E_\mathrm{M}$            & 0.044    &  0.036   &   0.033   &  0.054  \\
HFT apex                  & 1.94\%   &  0.54\%  &   \dots   &  \dots \\
CFL apex                  & 4.23\%   &  0.63\%  &   2.70\%  & 0.38\% \\
$\langle\,\Phi\,\rangle$  & 2.14\%   &  1.10\%  &   2.69\%  & 0.44\% \\
$\tilde n  $              & 3.92\%   &  3.89\%  &   6.61\%  & 0.51\% \\
$E_\mathrm{mag}$          & 0.22\%   &  0.37\%  &   0.74\%  & 0.03\% \\
$H_\mathrm{mag}$          & 4.55\%   &  4.43\%  &   4.83\%  & 2.80\% \\
\hline
\end{tabular}
\flushleft
\footnotesize{Figures of merit in the analysis domain as defined in Sect.~\ref{s:results}, with smaller values indicating higher accuracy. Deviations from the corresponding values in the reference fields are given in percent. }
\end{table}

\subsection{Stable equilibria}\label{s:bavshft}

We consider the reconstruction of four stable TD equlibria, two of which contain an HFT (of different height) and one contains a BP (Table~\ref{t:tdparam}), using the potential field by \cite{1964NASSP..50..107S} as initial condition. 
The main results are summarized in Table~\ref{tb:potvgerr}, where measures of the reconstruction quality of the four reference fields are listed (see also the comprehensive compilation of the various metrics for all our extrapolations in Table~\ref{t:tddd}). 
The table shows that the errors are very small and of comparable magnitude for all cases (within fractions of a percent up to a few percent). 
In particular, we point out that the mean vector error $E_\mathrm{M}$ of all four reconstructed fields is 0.054 or smaller, a result \emph{far} better than the accuracy achieved in \emph{any} previous similar reconstruction study, i.e., in those that used only the vector magnetogram as input \citep{2006SoPh..235..161S, 2006A&A...453..737W, 2007SoPh..245..263V, 2008SoPh..247..269M}, although a smaller margin between the analysis and extrapolation volumes is used here. 
This mean vector error corresponds to a mean angular deviation between $\vB$ and $\vB^\mathrm{ref}$ of at most 3~degrees (if the field strengths agree on all grid nodes) and to a mean deviation of field strength of at most 5\% (if all directions agree)---values lower than the noise level for most pixels in observed magnetograms. 
It is also worth noting that the reconstruction accuracy is now comparable to that achieved previously in ``extrapolations'' that have provided the reference field vector on all six boundaries of the reconstruction volume. 
The accuracy reached here is clearly superior to all but one of those results reported in Tables~1 and 2 in \citet{2006SoPh..235..161S} and is approaching the best one in this paper, as well as the results in \citet{2006A&A...446..691A} and \citet{2006A&A...453..737W}. 
The field line plots in Fig.~\ref{f:fl_ex14} show that such small errors have practically no visible influence on the reconstruction, which is virtually indistinguishable from the reference field. 

The best reconstruction is obtained for the case with a bald patch, while the least accurate one is the No\_HFT case, where neither an HFT nor a BP are present. 
There is apparently no relation between the quality of the reconstruction (Table~\ref{tb:potvgerr}) and the compatibility of the corresponding magnetogram with the force-free constraints (Table~\ref{t:mgmff}). 
This shows that the level of inconsistencies in the vector magnetograms of our stable reference fields is sufficiently small to have essentially no influence on the extrapolation quality. 

\begin{figure}
 \resizebox{\hsize}{!}{\includegraphics{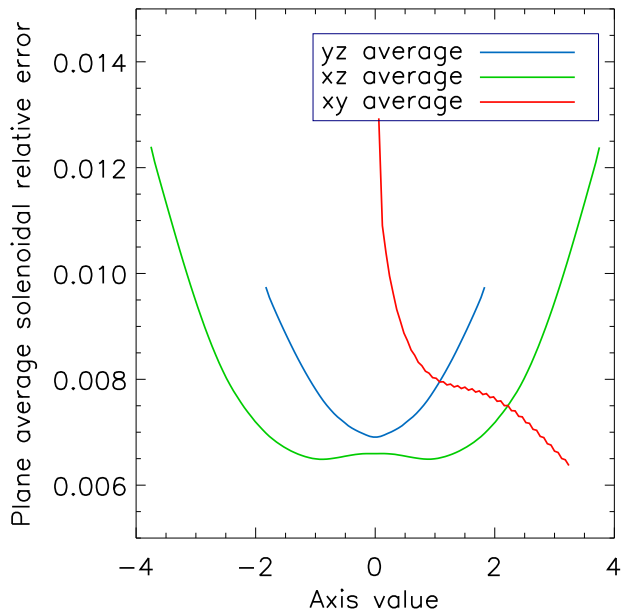}}
 \caption{Plane averages of the relative solenoidal error $|\vB_\mathrm{ns}|/|\vB|$ in the No\_HFT case, as a function of position.\label{f:av_bns}}
\end{figure}
Turning to a more detailed evaluation, we first consider the consistency measures of the reconstructed fields given in the upper part of Table~\ref{tb:potvgerr}. 
On average, the angle between the magnetic field and the current density stays below 1~degree, implying that the reconstructed fields have reached a high degree of force-freeness. 
The errors in satisfying $\divb=0$ as measured by $\langle\,|f_i|\,\rangle$ are also very small, well below those of the potential field used as initial condition (which is nominally divergence-free, except for discretization and truncation errors).  
Similarly, $|\vB_\mathrm{ns}|/|\vB|$ is small. 
Let us consider this quantity for the No\_HFT case which has the highest value of $\langle\,|f_i|\,\rangle$. 
Since the directions of $\vB_\mathrm{ns}(\vec{x})$ are essentially random, a very conservative estimate of its influence is $\langle\,|\vB_\mathrm{ns}|/|\vB|\,\rangle$, which stays below $6\%$, with a grid average of $0.7\%$. 
Averaging the relative error $|\vB_\mathrm{ns}|/|\vB|$ in the planes of constant $x$, $y$, and $z$, yields the profiles plotted in Fig.~\ref{f:av_bns}. 
These show that the deviations from the solenoidal condition originate primarily at the boundaries of the box, actually being smallest in the current channel in its interior.
Overall, the values of $\sj$, $\langle\,|f_i|\,\rangle$, and $\vB_\mathrm{ns}/|\vB|$ demonstrate that the reconstructed stable TD equilibria possess a high degree of consistency with the assumptions of the extrapolation method. 

The geometry of the reference field is perfectly recovered, as expected from the very low value of $E_\mathrm{M}$ and illustrated in Fig.~\ref{f:fl_ex14} for the case Low\_HFT. 
The flux rope, as the entire extrapolated field, reproduces the $z$-axis line symmetry of the TD equilibrium exactly. 
Therefore, the flux rope location is correctly represented by the apex height of the CFL, found to match the height in the reference field nearly exactly for all stable equilibria. 
Since $E_\mathrm{M}$ includes a normalization to the local field value on the grid, it gives the weak-field areas in the outer parts of the reconstruction volume the same weight as the strong-field areas near the bottom and in the flux rope. 
From the very small value of $E_\mathrm{M}$, we can therefore be sure that the high overlying field loops are reconstructed with essentially the same accuracy as the flux rope, and this is apparent from Fig.~\ref{f:fl_ex14} as well. 
The comparison of such loops with soft X-ray or EUV loops has proven insatisfactory in recent nonlinear extrapolations of observed vector magnetograms \citep{2009ApJ...696.1780D}, although the observed loops very likely outline force-free or even current-free field line bundles. 
The results given here demonstrate that this is not due to limitations of the extrapolation scheme(s). 

\begin{figure}
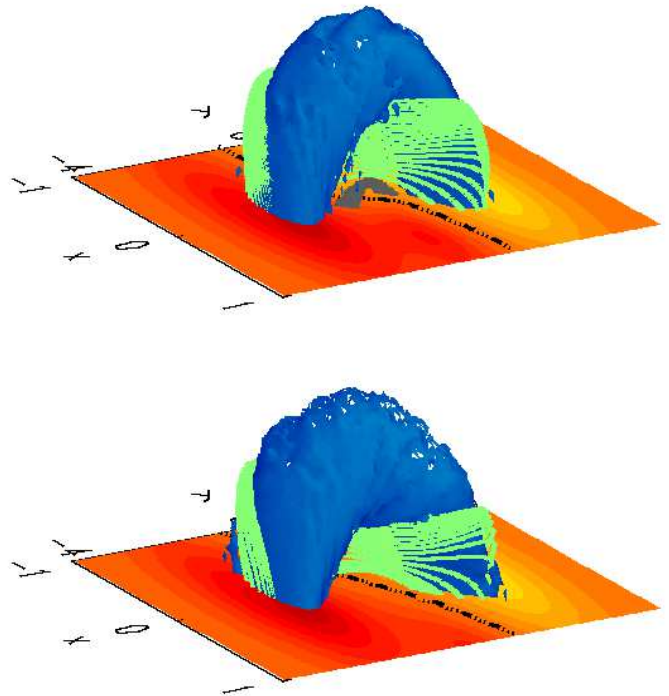

 \resizebox{\hsize}{!}{\includegraphics[bb=101 368 517 602, clip]{hft_ex15.eps}}\\
 \resizebox{\hsize}{!}{\includegraphics[bb=101 368 517 602, clip]{hft_ex16.eps}}
 \caption{Isosurface of the reconstructed current channel at 31\% of peak current density for the cases High\_HFT \emph{(top)} and No\_HFT \emph{(bottom)}. Field lines crossing the $z$ axis are shown in grey if below the HFT and in green if above the HFT.\label{f:hft}}
\end{figure}
The HFT is recovered for both reference fields containg one, with high accuracy of the HFT apex height. 
This is illustrated for the High\_HFT case in Fig.~\ref{f:hft}, which also includes the No\_HFT case for comparison. 
The bald patch in the BP case is recovered as well, occupying a similar section of the neutral line as in the reference field, see Fig.~\ref{f:isobp}. 
These results demonstrate that our magnetofrictional code is able to reconstruct the geometry and topology of TD-like equilibria.

Considering the parameters that characterize the relevant instabilities, we find that the relatively high twist of $\approx(1.8\mbox{--}2.1)\pi$ is recovered with very high accuracy, the errors being below 3\% in all four cases. 
Thus, as in \cite{2005A&A...433..335V}, it is confirmed that the magnetofrictional method can reconstruct field lines that wind around a flux rope axis more than one time.
In this regard the nonlinear extrapolation differs in principle from linear extrapolations, which limit the number of field line turns in the computed field to about 1/2 for realistic magnetogram sizes, due to the upper limit on the force-free parameter inherent in linear fields \cite[Kliem et al., in preparation; see also][]{2005ApJ...626.1091L}. 
This means that estimates of twist in fields extrapolated with the magnetofrictional code are by far more reliable than estimates using a linear code. 

\begin{figure}
 \resizebox{\hsize}{!}{\includegraphics[bb=36 21 605 586, clip]{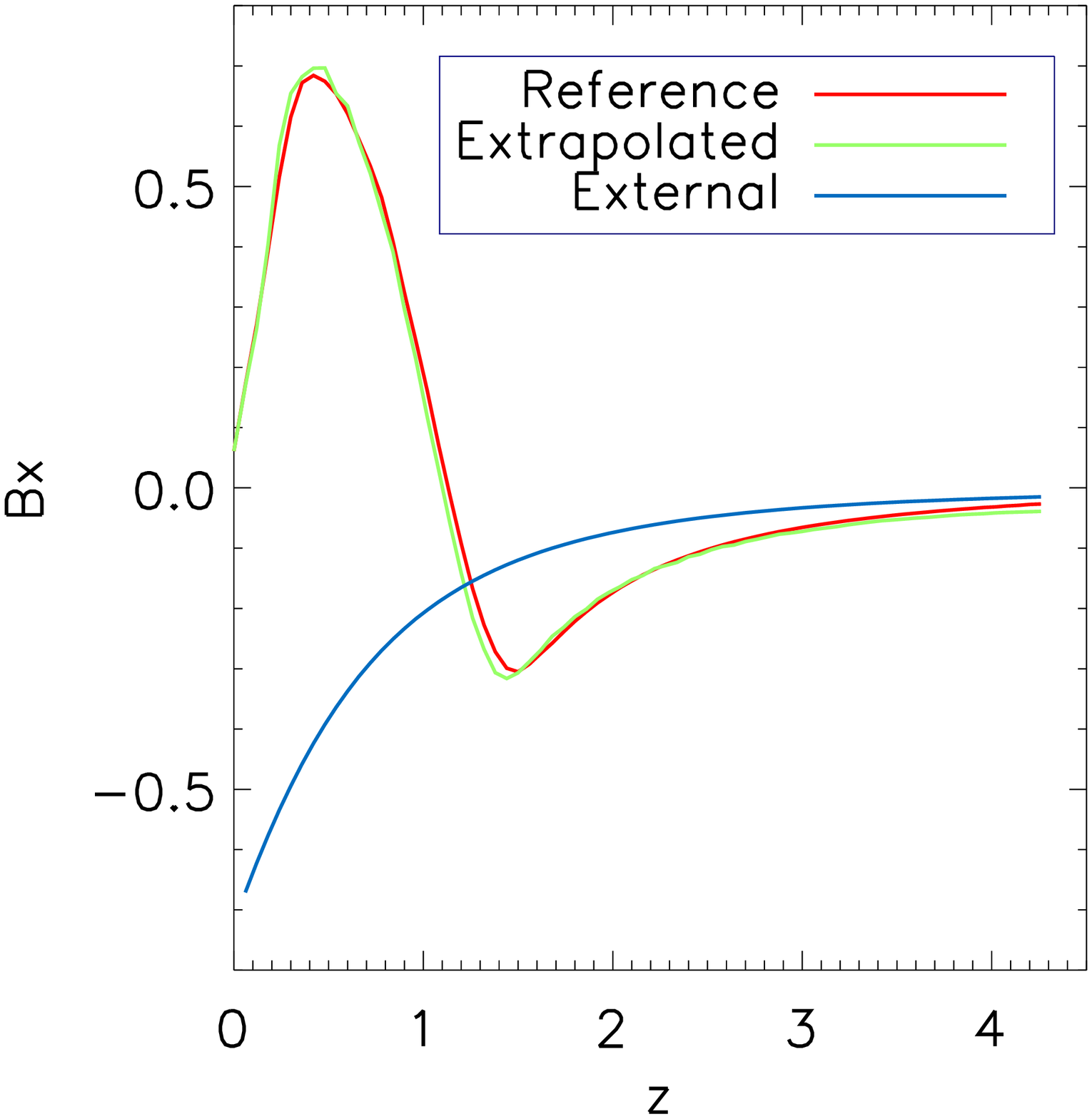}\includegraphics[bb=36 21 605 586,clip]{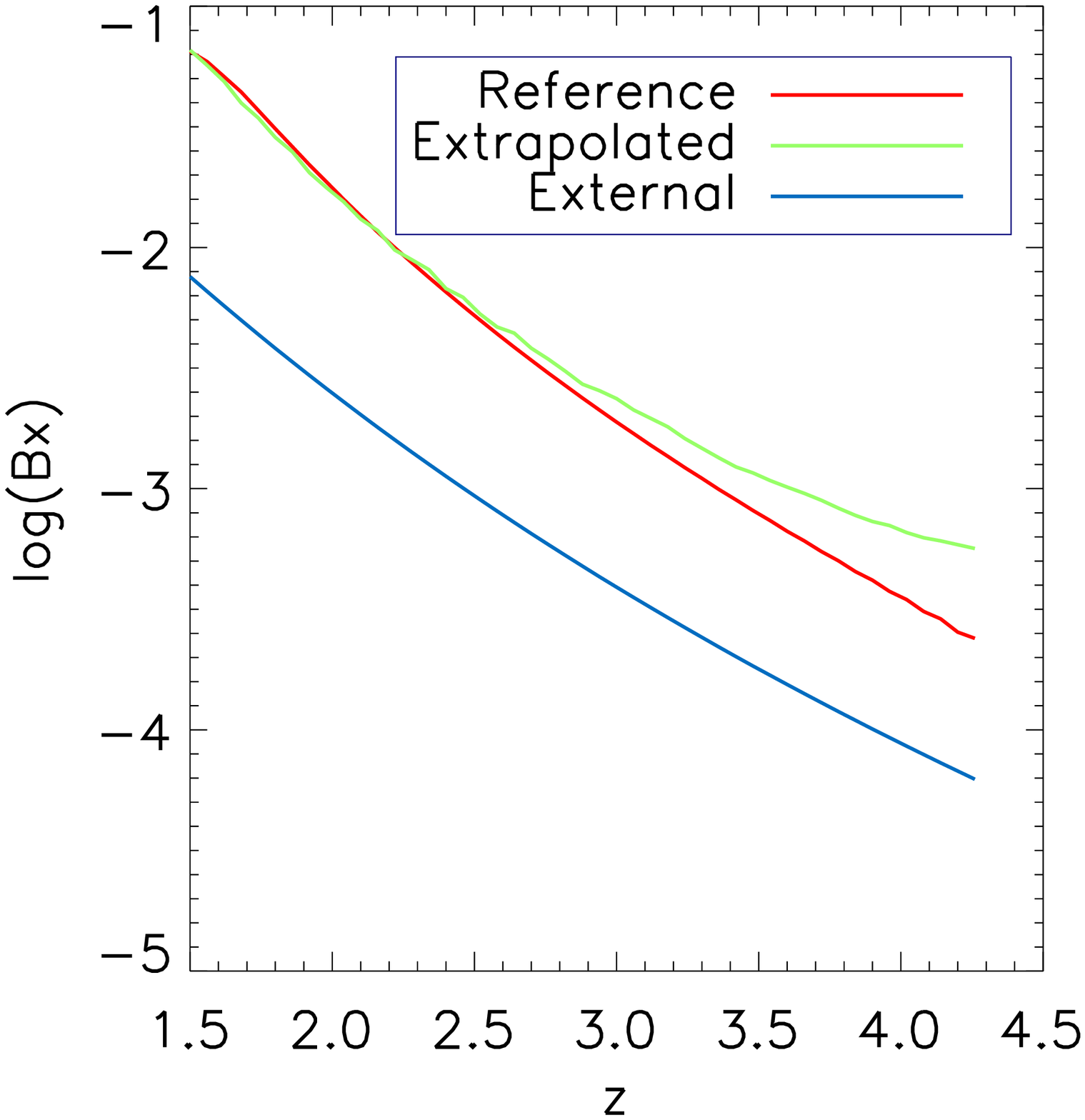}}
 \caption{Poloidal field at the $z$ axis, $B_x(0,0,z)$, in the No\_HFT case, for the reference (red), extrapolated (green), and TD\_ex potential fields (blue). 
\emph{Left}: entire $z$ axis extension; \emph{right}: upper part (above the current channel) in logarithmic scale.
Note that the offset of the blue curve results from normalizing the external field in the same manner as the full TD field, i.e., by the value $|B(0,0,1)|$, even though it does not contain a flux rope. \label{f:cfr_bx_z}}
\end{figure}
In order to assess the reliability of the extrapolation with regard to the torus instability, we consider the approximate decay index of the total poloidal field, $\tilde n$, given by Eq.~(\ref{eq:tilden}). 
The No\_HFT case is illustrated in Fig.~\ref{f:cfr_bx_z} to show that the reference and reconstructed profiles of $B_x(0,0,z)$ agree nearly exactly in the whole range of $z$ values. 
The same is true, of course, for the slope $\tilde n$, especially in the relevant region immediately above the current channel, as the logarithmic plot on the right-hand side of the figure shows. 
The very good agreement of the field profile at the whole $z$ axis implies that both parts of the poloidal field must be well recovered individually and confirms the use of $\tilde n$ as proxy for the reconstruction quality of $n$. 
Figure~\ref{f:cfr_bx_z} suggests the height range above the current channel not much exceeding $z=2$ for the computation of a representative value for $\tilde n$. 
We will use the range $z\in[1.80, 2.16]$. 
The values thus obtained differ by only 0.5--7\% from the true ones (Table~\ref{tb:potvgerr}), a precision better than that of the current knowledge of the torus instability threshold. 
The comparison with the corresponding profile of the external field components in the TD equilibrium (blue curves) illustrates that the slopes of the field, $n$ and $\tilde n$, indeed differ considerably in the height range used for the $\tilde n$ computation, so that the absolute values of this parameter, included in Table~\ref{t:tddd} for completeness, have no direct bearing on the torus instability. 
Since the close agreement of $\tilde n$ for the reconstructed and reference fields implies that $n$ is recovered with similar accuracy, we conclude that the key parameters which determine the onset of the two relevant instabilities are reliably reproduced. 

Similarly, the magnetic energy and relative magnetic helicity are obtained with high accuracy; their errors remain below 1 and 5 percent, respectively, for all four stable reference fields. 

Figure~\ref{f:fl_ex14} also shows that the potential and linear extrapolations entirely miss the flux rope, typically leading to an arcade-type structure.
Despite the failure of the potential field to reproduce any of the salient features of the reference field, still some of the comparison metrics score decently (\eg $E_\mathrm{M}=0.28$; see Table~\ref{t:tddd} for more detail). 
This is due to the fact that a large fraction of the analysis volume is occupied by potential field.
The energy of the potential field reconstruction, being the minimum for a given normal field distribution at the boundaries, is of course lower than the energy of the reference field.
However, it still accounts for a large part of it, about 80\% in the selected analysis volume.
The linear reconstruction suffers from similar limitations but, since it fills the entire volume with currents, it is affected by even larger errors than the potential field. 
The value $\alpha=-0.2$ adopted for the computation of the linear field is the one that gives the best r.m.s.\ matching of the transverse magnetogram components, often referred to as $\alpha_\mathrm{best}$ \cite[see][]{1995ApJ...440L.109P}.

Finally, we briefly comment on the reconstruction of the TD equilibrium in \cite{2006A&A...453..737W}. 
The original TD equilibrium with a line current was considered, and a different extrapolation scheme, the optimization method, was used. 
The paper includes a reconstruction based only on the information from the bottom boundary as ``Case~II.'' 
The corresponding field line plot in Fig.~2 of that article (done by one of the authors of the present paper, GV) indicates that the extrapolated field contains two partially twisted flux tubes rather than one.   
That plot is actually misleading because the appearance of two separate flux tubes has been obtained by calculating field lines starting around the nominal footpoint positions of the geometrical torus axis; however, as described above, the footpoints of the flux rope's magnetic axis do not coincide with those of the geometrical torus axis. 
A proper field line plot (not included here) is obtained by placing the field line start points around the magnetic axis, and it shows a single, higher and slightly writhed flux rope, similar to the cases analyzed here. 
Due to the many differences between the employed test fields and reconstruction volumes, we do not further pursue the comparison with the results in \cite{2006A&A...453..737W}. 
Those are far less accurate than the ones given here in the case that only the vector magnetogram is used as input.

\subsection{Dependence on the initial field}\label{s:inicond}

In this section we describe the link between the potential or linear force-free field that is used as initial condition for the extrapolation and the quality of the obtained reconstruction. 
A key consideration here is 
the balance between the hoop force and the Lorentz force provided by the external poloidal field.
We consider the Low\_HFT case in combination with four different initial fields: two potential fields computed by the methods of \cite{1964NASSP..50..107S} and \cite{1978SoPh...58..215S}, respectively, a linear force-free field computed by the Seehafer method, and the external potential field of the TD equilibrium (generated by the magnetic charges and dipoles in the modified version of the equilibrium used in this paper). 
The latter, referred to as TD\_ex in the following, is, of course, not available for observed magnetograms. 

\begin{figure}
 \resizebox{\hsize}{!}{\includegraphics{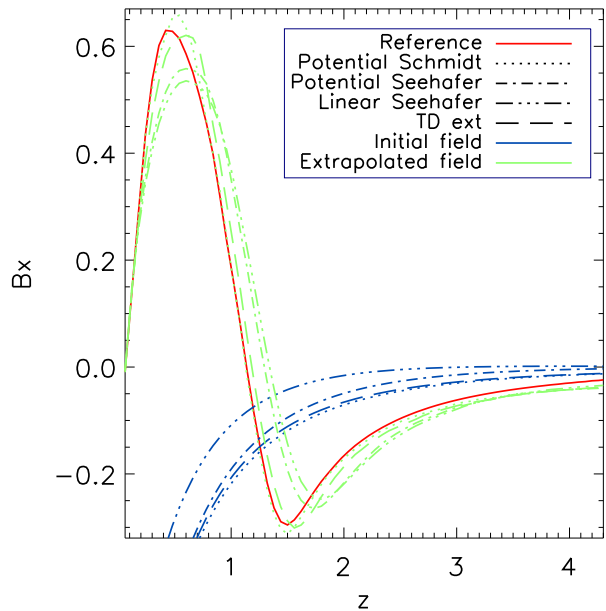}}
\caption{Poloidal field at the $z$ axis, $B_x(0,0,z)$, in the Low\_HFT case, for the reference (red solid line), initial (blue), and corresponding  extrapolated fields (green), for different type of the initial field: potential Schmidt (short dashes), potential Seehafer (dot--dash), linear alphabest Seehafer (triple dots--dash), and potential TD\_ex (long dashes).\label{f:cfr_bx_z_ini}} 
\end{figure}
Figure~\ref{f:cfr_bx_z_ini} shows the poloidal component at the $z$ axis for the reference field (red line), for the four fields used as initial condition (blue lines), and for the corresponding four reconstructions (green lines). 
The Schmidt and Seehafer potential fields are, of course, computed using the same $B_z(x,y,0)$ as boundary condition, but they differ in their assumptions about the field outside the magnetogram area. 
While Schmidt's method assumes the field to vanish outside (and therefore requires flux balance within the magnetogram), Seehafer's method uses a periodic repetition of an extended magnetogram obtained by mirroring the original magnetogram about one corner (in this way ensuring the flux balance). 
The resulting poloidal component of the Schmidt potential field approximates the external poloidal field of the TD equilibrium rather well, being larger by only a factor 1.06 at the nominal position of the CFL apex, $z=1.14$. 
Hence, the flux rope to be built by the extrapolation is provided with almost the correct external poloidal field, so that the force balance is indeed found very close to the correct position (Table~\ref{tb:lowxvgerr}).

\begin{table}
\caption{Reconstruction accuracy of the Low\_HFT equilibrium for different initial conditions}
\label{tb:lowxvgerr}
\centering
\begin{tabular}{ccccc}
\hline \hline
{Figure}  & {Potential} &          & {Potential} & {Linear}        \\
{of merit}& {Schmidt}   & {TD\_ex} & {Seehafer}  & {Seehafer}      \\
\hline
$\sigma_J \times 10^{2} $ & 0.73    &  1.18    &  4.12    &  4.52     \\
$\langle\,|f_i|\,\rangle \times 10^{5}$ & 6.90    & 11.98    & 16.18    & 14.47     \\
$E_\mathrm{M}$            & 0.036   &  0.072   &  0.065   &  0.070    \\
HFT apex                  & 0.54 \% &  1.05 \% &  1.77 \% &  2.06 \%  \\
CFL apex                  & 0.63 \% &  4.11 \% & 13.93 \% & 17.99 \%  \\
$\Phi$                    & 1.10 \% &  2.58 \% &  2.26 \% &  3.01 \%  \\
$\tilde n  $              & 3.89 \% &  2.45 \% & 17.54 \% & 36.88 \%  \\
$E_\mathrm{mag}$          & 0.37 \% &  1.49 \% &  2.30 \% &  2.26 \%  \\
$H_\mathrm{mag}$          & 4.43 \% &  7.58 \% &  1.38 \% &  6.86 \%  \\
\hline
\end{tabular}
\end{table}
The Seehafer potential and linear fields are weaker at the nominal CFL apex position, with a ratio of the poloidal components to $B_\mathrm{ep}$ of the TD\_ex field of 0.9 and 0.48, respectively, so that the flux rope finds equilibrium at greater heights. 
However, the apex heights differ by only 14\% and 18\%, respectively, which shows that the initial field does have an influence on the extrapolation, but also that the magnetofrictional relaxation can tolerate a substantial mismatch between the initial field and the true external field while still producing relatively small errors. 
The information flowing from the magnetogram into the extrapolation volume can largely, but not completely, modify the properties of the initial field, although open boundaries are implemented at the top and sides of the box, which allow the field to change in response to changes in the interior. 
This has to do with the fact that large parts of the volume are current free, both in the initial and in the final configuration. 
Hence, currents must be built up and subsequently be removed, in order to modify the initial field in the large outer parts of the box. 
This is a slow process, eventually hindered by numerical diffusion. 

Apart from the CFL apex height and the related decay index $\tilde{n}$, the figures of merit of the four extrapolations in Table~\ref{tb:lowxvgerr} show excellent values. 
The tendency for the Seehafer potential and linear fields to score worse is visible but rather weak, indicating that the initial field has a weaker influence on quantities like flux rope twist and total energy and helicity. 
These depend primarily on the information contained in the magnetogram and only to a smaller extent on the flux rope height. 
It is also clear that the figures of merit for the Schmidt potential field mostly beat those of TD\_ex because the former is more consistent with the magnetogram, which includes the contribution by the current channel. 

All four extrapolations recover the very low lying HFT apex of the test field, essentially at the correct height (Table~\ref{t:tddd}). 

These comparisons clearly suggest to use the Schmidt potential field as initial condition for flux-balanced magnetograms of isolated active regions (two conditions that go together in tendency). 
A flux imbalance can often be reduced by embedding the vector magnetogram in a larger magnetogram, which is typically of lower resolution and (with current instrumentation) only a line-of-sight magnetogram. 
The transverse components must then be modeled, for example by assuming a potential field, and this can strongly influence the outcome of the extrapolation. 
Alternatively, the use of the imbalanced vector magnetogram joint with open boundary conditions that allow for flux and currents to leave the numerical box through the top and side boundaries, as in our code, can lead to better extrapolations (see \citeauthor{2009ApJ...696.1780D} \citeyear{2009ApJ...696.1780D} and Fuhrmann et al.\ 2010 [in preparation] for a discussion). 
Finding the best trade-off between these options will require further study. 

\subsection{Unstable equilibrium}\label{s:unstable}

As mentioned in the Introduction, the parameters defining the TD equilibrium can be chosen such that the configuration is unstable with respect to the helical kink or to the torus instability.
Such equilibria offer the opportunity to test the extrapolation code in the reconstruction of unstable force-free configurations, addressing the question which signatures of the unstable situation are produced. 
We compare the extrapolation of the magnetogram provided by an unstable TD equilibrium with the equilibrium itself and with the post-eruption configuration obtained by a numerical MHD evolution of the perturbed equlibrium with the magnetogram kept fixed.

The configuration we consider is the same as the eruptive case in \citet{2005ApJ...630L..97T}, except that a smaller initial twist of $2.7 \pi$ is used (note that the twist average over the whole cross section of the current channel, as calculated by \citeauthor{2005ApJ...630L..97T}, is $\approx4\pi$ for this configuration). 
The MHD simulation is performed in a Cartesian box of size $[-20, 20]\times[-20, 20]\times[0, 40]$, which consists of the uniform grid used for the extrapolation in its central part and a streched grid outside the extrapolation domain.  
The smaller twist allows us to relax the configuration, including the magnetogram plane, for a few Alfv\'en times, reducing the initial spurious Lorentz force densities, before the helical displacement due to the developing kink instability becomes visible. 

\begin{figure}
 \resizebox{\hsize}{!}{\includegraphics[clip]{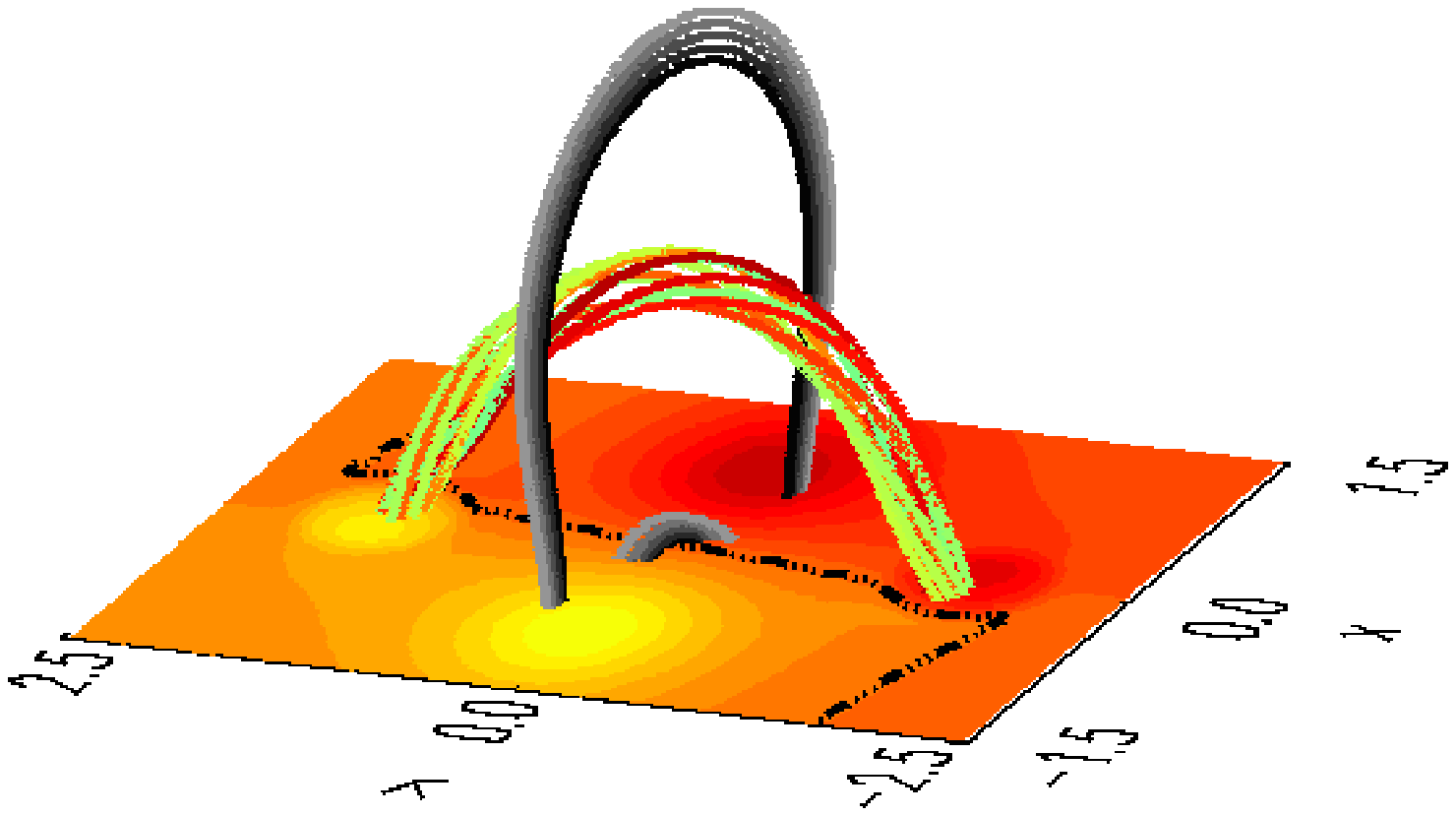}}\\
 \resizebox{\hsize}{!}{\includegraphics[clip]{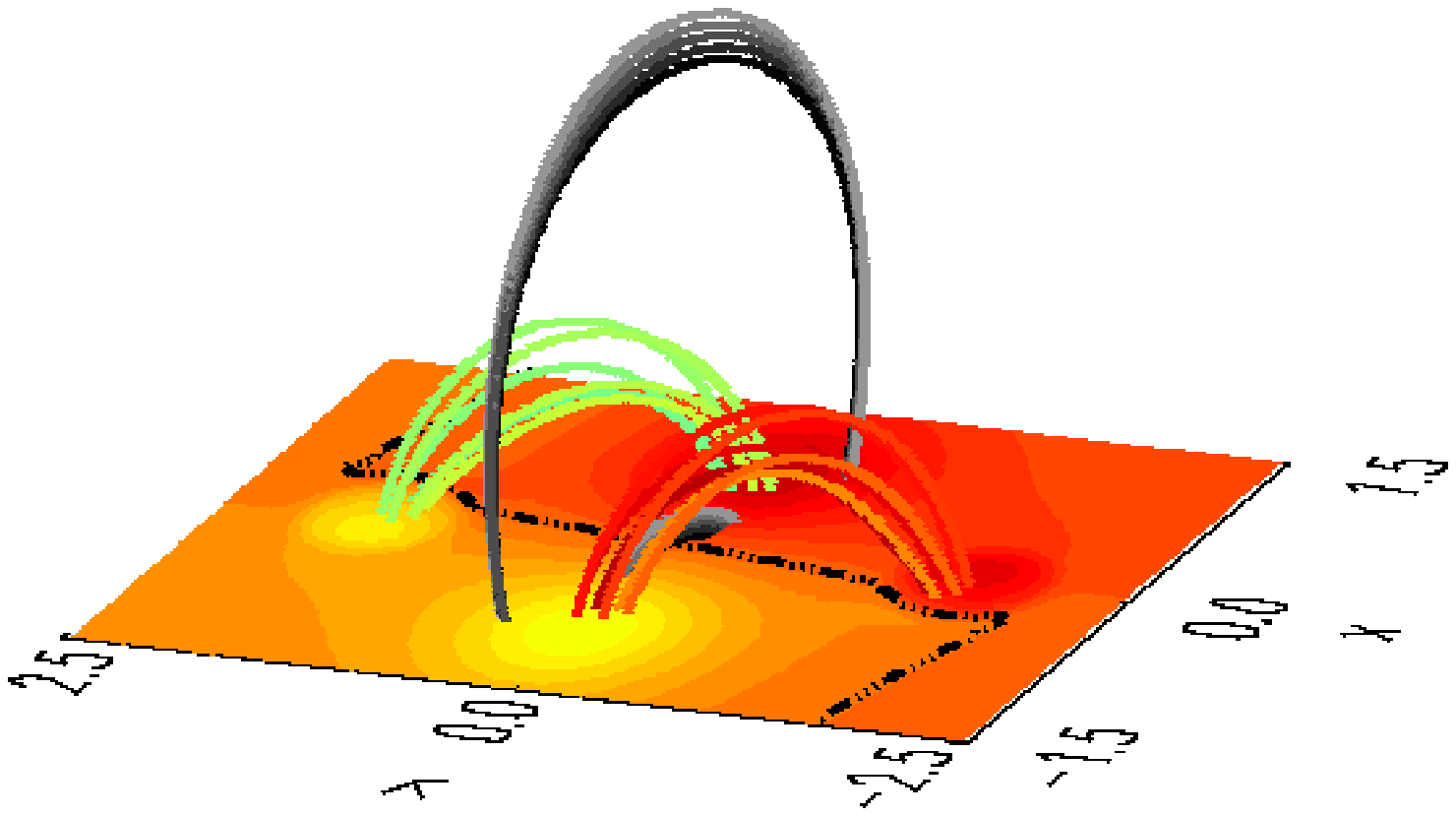}}
 \caption{Selected field lines for the reference fields of the unstable case at the initial \emph{(top)} and final \emph{(bottom)} stages of the MHD evolution (see Fig.~\ref{f:fl_ex14} for the color coding of the field lines).  \label{f:fl_erup_nl}}
\end{figure}
After this initial relaxation, we fix the magnetic field in $\{z=0\}$ for the whole subsequent MHD evolution. 
We refer to this partially relaxed, unstable configuration as the \emph{initial reference field} for the extrapolation of the unstable case; a field line plot is provided in Fig.~\ref{f:fl_erup_nl}. 
A small, transient, upward directed velocity perturbation is then applied at the flux rope apex to ensure that the rope kinks upwards. 
Once the kink instability has lifted the rope to a height where the field decreases sufficiently fast, the torus instability sets in.  
The rope is then additionally accelerated by the torus instability and erupts fully. 
In the wake of the rising flux rope, a vertical current sheet is formed. 
Reconnection in this current sheet cuts the legs of the flux rope at low heights (whereas the bulk of the rope still erupts), and the legs then connect to the sunspots. 
The upper part of the reconnected flux rope reaches the closed top boundary at $\sim90$ Alfv\'en times and is subsequently compressed there. 
In the remaining time of the simulation, which is terminated after 750 Alfv\'en times, changes in the lower part of the box occur mainly in the cusp-shaped arcade below the reconnection region, whereas the reconnected flux rope legs remain practically unchanged.

In order to be compatible with fixing the magnetogram, we set the plasma velocities to zero in the grid layers $\{z=0\}$ and $\{z=\Delta\}$. 
Therefore, although the simulation is run for a long time, the Lorentz force densities do not relax completely, so that the post-eruption configuration, although largely relaxed, is not perfectly force-free in the lower layers of the box. 
Nevertheless, the field line connectivities do not change significantly anymore, so that a qualitative comparison with the force-free extrapolation is possible.
We refer to this post-eruption configuration as the \emph{final reference field} for the extrapolation of the unstable case (see again Fig.~\ref{f:fl_erup_nl}).

The extrapolation of the unstable case is obtained in a similar way as described in Sects.~\ref{s:bavshft} and \ref{s:inicond}, except that here the extrapolation grid discretizes a larger volume $[-3.00, 3.00]\times[-5.04, 5.04]\times[ -0.06, 6.12]$ with the same uniform resolution $\Delta=0.06$. 
The extrapolation employs the $z=0$ magnetogram of the initial/final reference field and starts from the Schmidt potential field in $\{z>0\}$. 
Table~\ref{t:mgmff} verifies that the magnetogram includes residual forces, but it has a lower value of the applied torque than the stable cases. 
According to the metrics in the table, the unstable near-equilibrium TD configuration obtained from the initial partial MHD relaxation (before the instabilities become dominant) provides a magnetogram that is indeed acceptably force free. 

\begin{figure}
  \resizebox{\hsize}{!}{\includegraphics[clip]{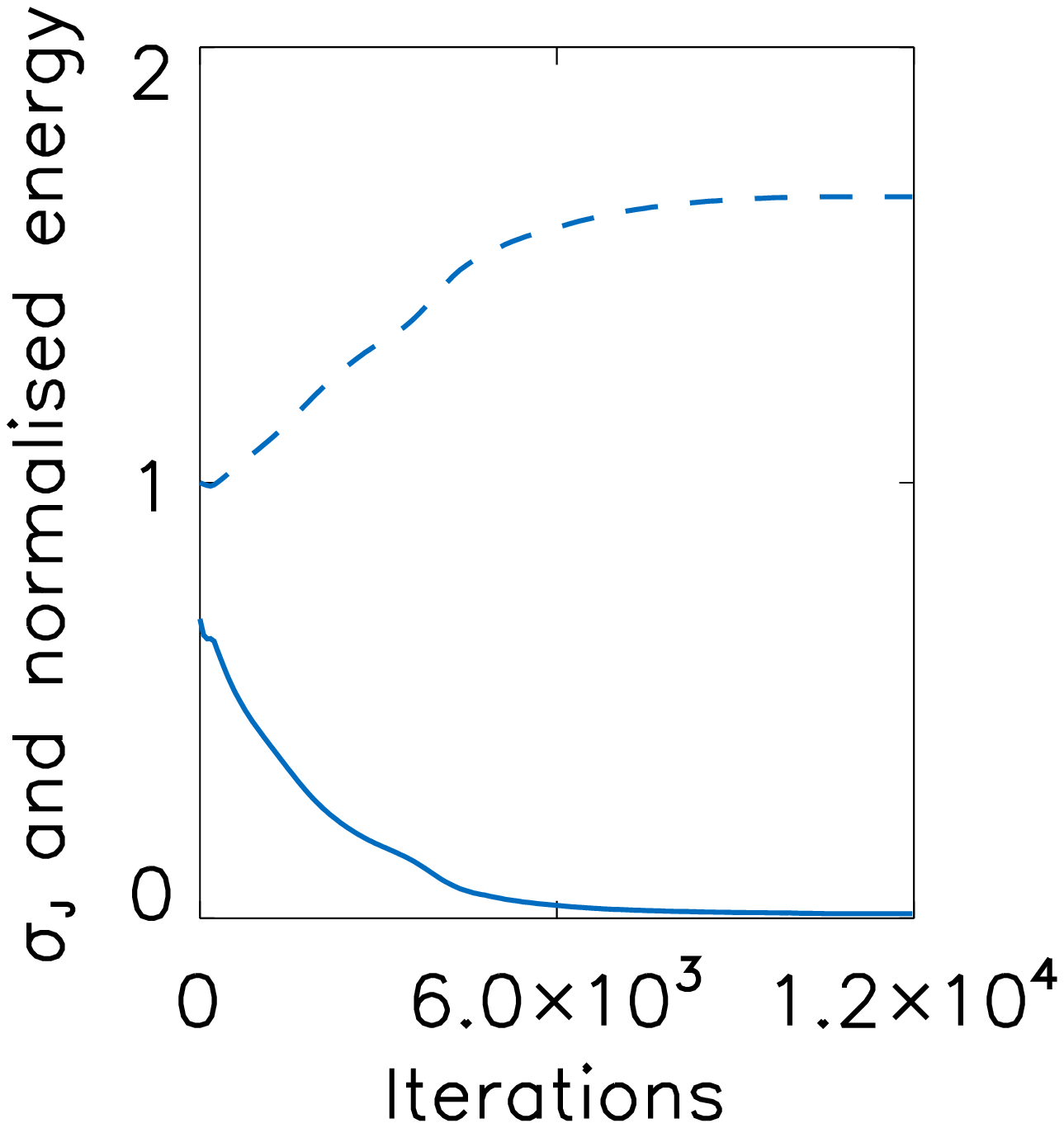}\includegraphics[clip]{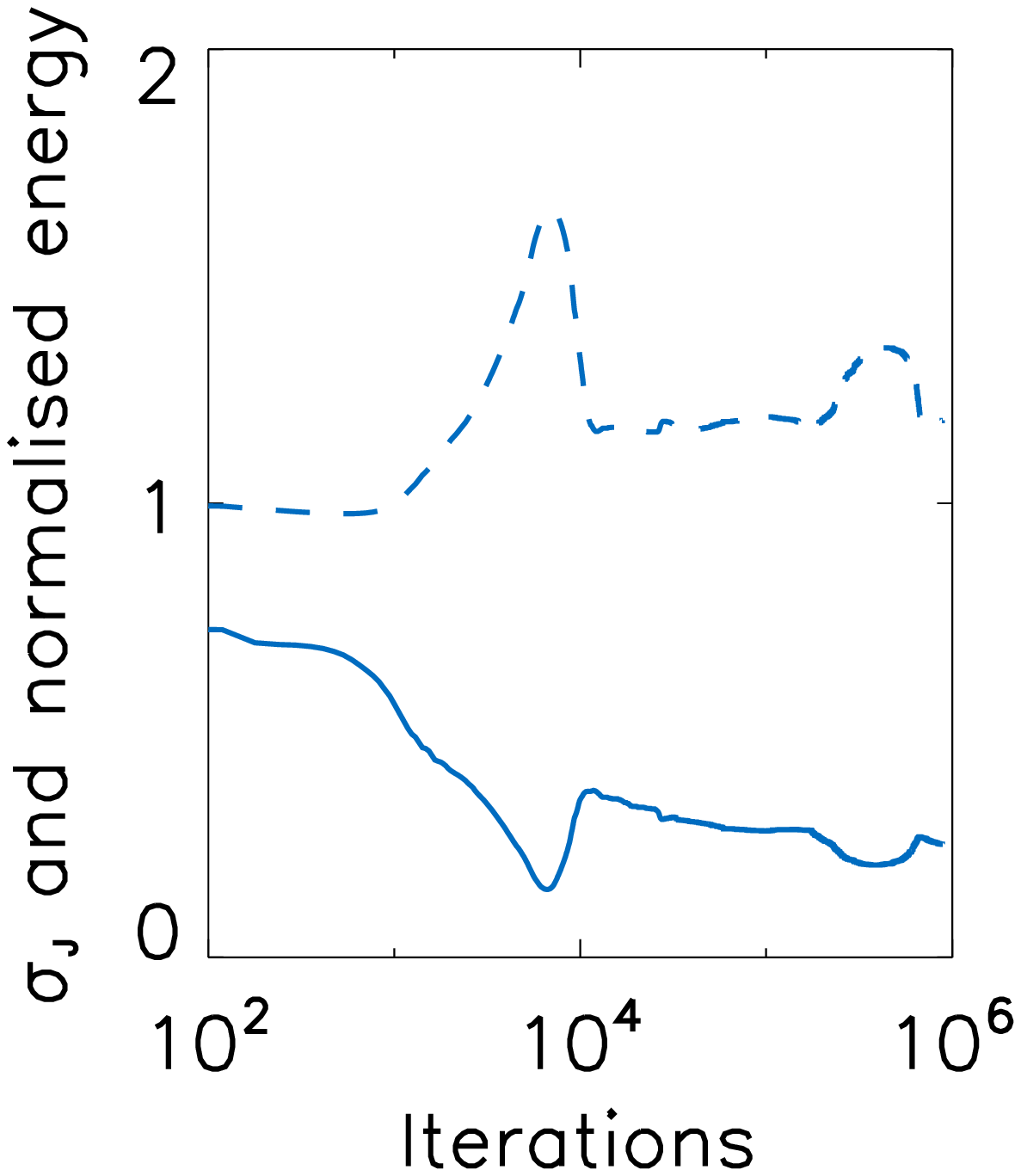}}
 \caption{Evolution of $\sj$ (solid line) and $E_\mathrm{mag}$ (dashes) during the extrapolation runs of the cases BP \emph{(left)} and Unstable \emph{(right)}. The energy is normalised to its value at the first iteration. \label{f:sj}}
\end{figure}
In all stable cases analyzed in the previous sections, the magnetic energy and $\sj$ evolve monotonically during the whole extrapolation run ($E_\mathrm{mag}$ increases, $\sj$ decreases), until nearly constant values are reached in $\sim7000$ iterations. 
The magnetic field hardly changes after this point, even if the magnetofrictional relaxation is continued for a very large number of iterations. 
An example of this behavior is given in Fig.~\ref{f:sj} for the BP case.
The extrapolation of the unstable case is significantly different. 
As Fig.~\ref{f:sj} shows, the energy (resp., $\sj$) reaches a temporary maximum (resp., minimum) in about the same number of iterations, but then the field changes and rapidly reaches an almost flat plateau with a lower energy and higher $\sj$.
This plateau extends over $\sim2\times10^5$ iterations, followed by a very gentle ``bump'' extending over $\sim5\times10^5$ iterations, and then by a second plateau. 
In the whole evolution from the beginning of the first plateau, the field line plots show that there is hardly any change in the magnetic configuration, so that we terminated the run after $9\times10^{5}$ iterations. 
From the extrapolation run we select a \emph{maximum-energy} and a \emph{minimum-energy} reconstruction, respectively corresponding to the maximum value of the energy at 6700 iterations and to the almost static state reached at the end of the run (referred to as ``MF Schmidt max'' and ``MF Schmidt min'' in Table~\ref{t:tddd}). 

\begin{figure*}
 \centering
  \includegraphics[width=17cm,clip]{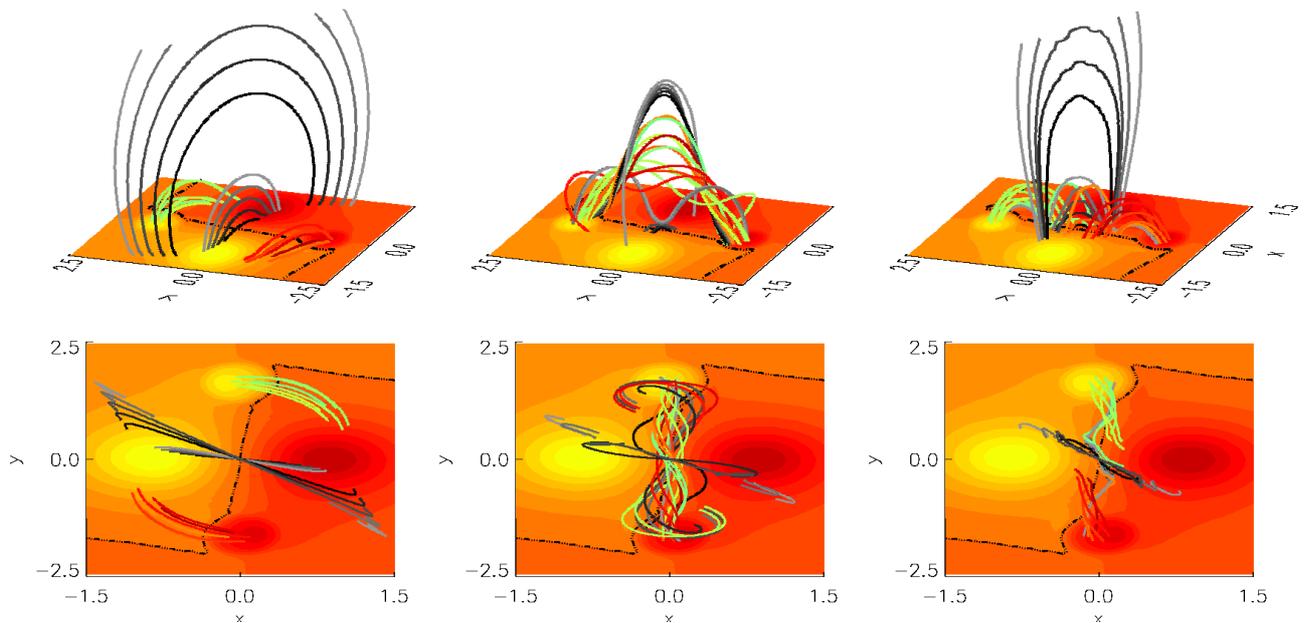}
 \caption{Selected field lines for the extrapolation of the unstable case at different iterations. From left to right: the initial potential field at it=1, the maximum-energy configuration at 6700 iterations, and the final low-energy configuration at $9\times 10^5$ iterations. See Fig.~\ref{f:fl_ex14} for the color coding of the field lines. \label{f:fl_erup_ex}}
\end{figure*}
The maximum-energy reconstruction is a relatively force-free ($\sj=0.13$) and divergence-free field, albeit the corresponding metrics reach only values that are about one order of magnitude worse than for the stable cases (see Table~\ref{t:tddd}).
Consequently, the r.h.s.\ of Eq.~(\ref{eq:mfintegral}) does not vanish, although $\partial_t E_\mathrm{mag}=0$, which illustrates the limitations of the applicability of Eq.~(\ref{eq:mfintegral}) discussed in Sect.~\ref{s:method}. 
This state is clearly a transitory one, as is the initial reference field. 
The two are determined by a transition between processes of partially different character, so that one can expect them to be neither very similar nor completely different. 
In the MHD run, the transition occurs between the relaxation of the initial Lorentz forces in a field that already includes a current channel and the dominant evolution of the instabilities, which release part of the free magnetic energy contained in the current channel. 
In the magnetofrictional run, the transition occurs between a phase of buildup of the current channel and the subsequent relaxation to a minimum-energy state, which destroys the channel. 
The maximum-energy field contains a flux rope with an HFT underneath, but this rope is not as compact as in the initial reference field (compare the top panel of Fig.~\ref{f:fl_erup_nl} with the middle panel in Fig.~\ref{f:fl_erup_ex}). 
Both the HFT and CFL reach about half the corresponding heights of the initial reference field.

We recall that the extrapolation starts from a potential field that has obviously no flux rope (as shown in the left panel of Fig.~\ref{f:fl_erup_ex}), and therefore the flux rope in the maximum-energy reconstruction is formed by the extrapolation process.
The extrapolation does not succeed to form the flux rope completely because the underlying TD configuration is in the unstable domain, so that the Lorentz forces tend to propagate the information about the current channel from the magnetogram into the volume and tend to move the system away from that configuration at the same time. 
The partial buildup of the current channel is also reflected by the fact that the maximum-energy field has reached 77\% of the energy in the initial reference field. 
In light of this number, the very close agreement of the magnetic helicities must be considered a coincidence or a consequence of the approximate nature of its calculation \cite[using the expressions in][]{2000ApJ...539..944D}. 

Conversely, the minimum-energy solution has no flux rope: field lines started at the original location of the flux rope connect to the sunspots (rather than to the flux rope footpoints in the TD and initial reference fields), in a manner largely similar to the final reference field (compare the bottom panel in Figs.~\ref{f:fl_erup_nl} with the right panel in \ref{f:fl_erup_ex}). 
The magnetic energy has a smaller relative error (of 16\%) than the maximum-energy reconstruction, while $E_\mathrm{M}=0.37$ is a rather poor match. 
Although both the MHD and magnetofrictional runs eventually proceed to a low-energy state, the closed box 
used in the MHD run prevents the system from relaxing as deep as in the magnetofrictional run. 
Furthermore, as argued above, we cannot expect the reconstruction of the final reference field to be accurate, since this field presents relatively strong forces close to the magnetogram due to the MHD eruption process. 

\begin{table*}
\caption{Extrapolations results for all TD equilibria}
\label{t:tddd}
\centering
\begin{tabular}{cccccccccc}
\hline \hline
\bf{Field}  & {HFT apex} &  {CFL apex }   & {$\langle\,\Phi\,\rangle / \pi$}  & 
{$\tilde n$} & {$1-E_\mathrm{M}$} & {$\sigma_J \times 10^{2} $} & {$E_\mathrm{mag}$} & 
{$H_\mathrm{mag}$} & {$\langle\,|f_i|\,\rangle \times 10^{5}$} \\
\hline
\multicolumn{10}{c}{High\_HFT}\\
\hline
Reference    &  0.1458 &  1.143 & -2.146 &  2.519 &  1.0000 &  0.5062 &  10.095 &  3.253 &  4.679 \\ 
MF Schmidt   &  0.1430 &  1.191 & -2.192 &  2.617 &  0.9561 &  1.5481 &  10.117 &  3.105 &  6.044 \\
\hline
\multicolumn{10}{c}{Low\_HFT}\\
\hline
Reference    &  0.0635 &  1.137 & -2.128 &  2.435 &  1.0000 &  0.3789 &  10.087 &  3.197 &  4.303 \\
MF Schmidt   &  0.0639 &  1.130 & -2.121 &  2.340 &  0.9645 &  0.7333 &  10.050 &  3.055 &  6.899 \\
MF Pot Seehafer&0.0624 &  1.295 & -2.176 &  2.008 &  0.9355 &  4.1202 &  10.319 &  3.241 & 16.18 \\
MF Lin Seehafer&0.0622 &  1.341 & -2.192 &  1.537 &  0.9296 &  4.5231 &  10.315 &  3.416 & 14.47 \\
MF TD\_ex    &  0.0629 &  1.183 & -2.183 &  2.494 &  0.9284 & 1.1770 &  10.237 &  2.954 & 11.98 \\
\hline
\multicolumn{10}{c}{Low\_HFT initial field}\\
\hline
    Schmidt  &  \dots &  \dots & \dots&  1.928 &  0.7246 & \dots &   7.958 &  0        & 277.9 \\
Pot Seehafer &  \dots &  \dots & \dots&  2.512 &  0.6278 & \dots &   8.143 &  0        &  1.247 \\
Lin Seehafer &  \dots &  \dots & \dots&  4.130 &  0.6665 & \dots &   8.267 &  2.880    &  1.225 \\
\hline
\multicolumn{10}{c}{No\_HFT}\\
\hline
Reference    &  \dots &  1.131 & -2.111 &  2.350 &  1.0000 &  0.2951 & 10.047  &  3.105 &  4.254 \\
MF Schmidt   &  \dots &  1.101 & -2.097 &  2.195 &  0.9670 &  1.1756 &  9.972  &  2.955 & 10.28 \\
\hline
\multicolumn{10}{c}{BP}\\
\hline
Reference    &  \dots &  1.389 &  -1.806 & -0.6210 & 1.0000 &  0.0538 &  7.489  &  3.610 &  7.162 \\
MF Schmidt   &  \dots &  1.385 &  -1.814 & -0.6242 & 0.9460 &  0.8003 &  7.487  &  3.509 & 10.05 \\
\hline
\multicolumn{10}{c}{Unstable}\\
\hline
Reference ini&  0.1705 &  1.008 & -2.853 &  2.916 &  1.0000 &  3.4610 &   3.286 & 24.21 &  3.841 \\
Reference fin&  \dots &  \dots & \dots&  2.410 &  1.0000 & 22.116 &   2.205 & 23.69 &16.86 \\
MF Schmidt max
             &  0.0984 &  0.4095&-4.115 &  1.992 &  0.7693 & 13.140 &   2.540 & 25.35 &94.96 \\
MF Schmidt min
             &  \dots &  \dots & \dots& \dots &  0.6266 & 27.396 &   1.843 & 23.08 &115.9\\
\hline
\end{tabular}
\flushleft
\footnotesize{
Reconstruction measures (absolute values) for all extrapolations presented in this paper, belonging to the cases No\_HFT, Low\_HFT, High\_HFT, BP, and Unstable, supplemented by similar analysis for the different initial fields used in the Low\_HFT case.
The leftmost column characterizes the analyzed field using the following naming convention:
`Reference' identifies the model field to be reconstructed by the extrapolation;
`MF' stands for nonlinear magnetofrictional extrapolation, `Pot' and `Lin' for potential and linear extrapolation, respectively;
`Schmidt' and `Seehafer' denote the two methods used to compute the initial field for the magnetofrictional extrapolations; `TD\_ex' is the external potential field generated by the magnetic sources and dipoles in the TD model. 
For the unstable case, the reference fields after the initial relaxation (Reference ini) and at the end (Reference fin) of the MHD simulation are analyzed.
Similarly, two extrapolated fields are reported, corresponding to the maximum (MF Schmidt max) and minimum (MF Schmidt min) energy of the reconstructed field (Sect.~\ref{s:unstable}).
All quantities are defined in  Sect.~\ref{s:results} and computed in the analysis domain. 
The derivatives involved in $\sigma_J$ and $\langle\,|f_i|\,\rangle$ are computed to fourth order accuracy, except for the reference fields and for the potential and linear fields, where a second order differencing is used, and one-sided derivatives of corresponding order are employed at, and where necessary also just above, the plane $\{z=0\}$. 
Missing values represent ill-defined quantities for the respective equilibria.
}
\end{table*}
\section{Conclusions}\label{s:conclusions}

This investigation demonstrates that a force-free model of solar active regions, which includes relevant structural (i.e., topological or geometrical) features and is free of noise, can be reconstructed to a very high accuracy based only on its vector magnetogram. 
The reconstructed configuration consists of a flux rope with a non-neutralized current channel in its core. 
A bald patch separatrix surface or two quasi-separatrix layers combined into a hyperbolic flux tube form the interface between the flux rope and the ambient potential field. 
These elements are regarded to be generic for many active regions, so that the model realistically captures their structural properties, in particular for essentially bipolar flux distributions.

Previous reconstructions of nonlinear force-free test fields considered the simpler configurations of an approximately current-neutralized flux rope \citep{2005A&A...433..335V} and of a sheared arcade \citep{2006SoPh..235..161S, 2007SoPh..245..263V}. 
A force-free model of a filament channel containing a hollow current channel could also be reconstructed, albeit at significantly lower accuracy, which was likely due to the combination with a noisy line-of-sight magnetogram in the exterior of the channel \citep{2008SoPh..247..269M}. 
Thus, the reconstruction capability has now been demonstrated, mostly at excellent accuracy, for all basic types of force-free equilibria currently considered in coronal physics, with the one treated here providing the highest degree of structural complexity. 
This confirms and substantiates previous conclusions \citep{2008SoPh..247..269M, 2009ApJ...696.1780D} that insufficient accuracy of nonlinear force-free extrapolation originates from the violation of the force-free assumption by the input vector magnetogram, not from insufficient intrinsic accuracy of the extrapolation codes. 

The accuracy achieved in the reconstruction of stable modified TD equilibria clearly permits to discriminate between configurations containing an HFT, a BP, or none of these features. 
The energy, flux rope twist and position, and helicity are obtained at relative error levels of $\la$\,1, 3, 4, and $\sim\!5\%$, respectively, even for a high twist near the threshold of the helical kink instability, slightly exceeding one full turn. 

These quantities are found to depend only weakly on the initial field chosen, except for the apex height of the flux rope which shows a displacement from the true value of up to $\approx20\%$ (if a linear force-free field is used). 
Starting the extrapolation from the conventional potential field, which assumes vanishing flux outside the magnetogram, yields the most accurate results for the considered equilibria. 
Other potential fields result in somewhat lower accuracy, and a linear force-free field scores worse but still produces a rather acceptable reconstruction. 

The high accuracy is reached using vector magnetograms that have a realistic low-flux margin of width $\sim1/4$ the size of the strong-flux area and are exactly flux balanced. 
The effects that result from violating the latter condition require further study. 

Attempting to extrapolate the magnetogram of an unstable modified TD equilibrium, the magnetofrictional relaxation produces an evolution in two phases of opposing tendency which clearly reflect the unstable nature of the equilibrium. 
A phase of rising free energy yielding a partial buildup of the flux rope is superseded by a decrease of the energy back to nearly the initial (potential-field) value, destroying the flux rope. 

\acknowledgements
We thank G. Aulanier and the referee for constructive comments.
This work was supported by 
the DFG, an STFC Rolling Grant, and by NASA grants NNH06AD58I and  NNX08AG44G. 
The research leading to these results has received funding
from the European Commission's Seventh Framework Programme
(FP7/2007-2013) under the grant agreement $n^\circ$ 218816 (SOTERIA
project, www.soteria-space.eu). Funding from the European
Comission through the SOLAIRE network (MTRM-CT-2006-035484)
is also gratefully acknowledged. 
The contribution of V.~S.~Titov was supported by NASA's Heliophysics Theory, Living
With a Star, and SR\&T programs, and the Center for Integrated Space Weather Modeling
(an NSF Science and Technology Center).

\bibliographystyle{aa}

\begin{thebibliography}{70}
\expandafter\ifx\csname natexlab\endcsname\relax\def\natexlab#1{#1}\fi

\bibitem[{{Alexiades} {et~al.}(1996){Alexiades}, {Amiez}, \&
  {Gremaud}}]{1994alexiades}
{Alexiades}, V., {Amiez}, G., \& {Gremaud}, P. 1996, in IMACS'94, vol.2

\bibitem[{{Amari} {et~al.}(1997){Amari}, {Aly}, {Luciani}, {Boulmezaoud}, \&
  {Mikic}}]{1997SoPh..174..129A}
{Amari}, T., {Aly}, J.~J., {Luciani}, J.~F., {Boulmezaoud}, T.~Z., \& {Mikic},
  Z. 1997, \solphys, 174, 129

\bibitem[{{Amari} {et~al.}(2006){Amari}, {Boulmezaoud}, \&
  {Aly}}]{2006A&A...446..691A}
{Amari}, T., {Boulmezaoud}, T.~Z., \& {Aly}, J.~J. 2006, \aap, 446, 691

\bibitem[{{Amari} {et~al.}(1999){Amari}, {Boulmezaoud}, \&
  {Mikic}}]{1999A&A...350.1051A}
{Amari}, T., {Boulmezaoud}, T.~Z., \& {Mikic}, Z. 1999, \aap, 350, 1051

\bibitem[{{Antiochos}(1998)}]{1998ApJ...502L.181A}
{Antiochos}, S.~K. 1998, \apjl, 502, L181+

\bibitem[{{Bungey} {et~al.}(1996){Bungey}, {Titov}, \&
  {Priest}}]{1996A&A...308..233B}
{Bungey}, T.~N., {Titov}, V.~S., \& {Priest}, E.~R. 1996, \aap, 308, 233

\bibitem[{{Canou} {et~al.}(2009){Canou}, {Amari}, {Bommier}, {Schmieder},
  {Aulanier}, \& {Li}}]{2009ApJ...693L..27C}
{Canou}, A., {Amari}, T., {Bommier}, V., {et~al.} 2009, \apjl, 693, L27

\bibitem[{{Chiu} \& {Hilton}(1977)}]{1977ApJ...212..873C}
{Chiu}, Y.~T. \& {Hilton}, H.~H. 1977, \apj, 212, 873

\bibitem[{{Chodura} \& {Schl{\"u}ter}(1981)}]{1981JCoPh..41...68C}
{Chodura}, R. \& {Schl{\"u}ter}, A. 1981, J.~Comp.~Phys., 41, 68

\bibitem[{{Craig} \& {Sneyd}(1986)}]{1986ApJ...311..451C}
{Craig}, I.~J.~D. \& {Sneyd}, A.~D. 1986, \apj, 311, 451

\bibitem[{{Demoulin} {et~al.}(1996){Demoulin}, {Henoux}, {Priest}, \&
  {Mandrini}}]{1996A&A...308..643D}
{Demoulin}, P., {Henoux}, J.~C., {Priest}, E.~R., \& {Mandrini}, C.~H. 1996,
  \aap, 308, 643

\bibitem[{{DeRosa} {et~al.}(2009){DeRosa}, {Schrijver}, {Barnes}, {Leka},
  {Lites}, {Aschwanden}, {Amari}, {Canou}, {McTiernan}, {R{\'e}gnier},
  {Thalmann}, {Valori}, {Wheatland}, {Wiegelmann}, {Cheung}, {Conlon},
  {Fuhrmann}, {Inhester}, \& {Tadesse}}]{2009ApJ...696.1780D}
{DeRosa}, M.~L., {Schrijver}, C.~J., {Barnes}, G., {et~al.} 2009, \apj, 696,
  1780

\bibitem[{{DeVore}(2000)}]{2000ApJ...539..944D}
{DeVore}, C.~R. 2000, \apj, 539, 944

\bibitem[{{Evje} {et~al.}(2000){Evje}, {Karlsen}, {Lie}, \&
  {Risebro}}]{2000IMA.120...19E}
{Evje}, S., {Karlsen}, K., {Lie}, K., \& {Risebro}, N. 2000, in {IMA Volumes in
  Mathematics and its Applications}, Vol. 120, {Parallel Solution of Partial
  Differential Equations}, ed. P.~{Bjorstad} \& M.~{Luskin} (Springer Science +
  Business Media, LLC), 209--228

\bibitem[{{Fuhrmann} {et~al.}(2007){Fuhrmann}, {Seehafer}, \&
  {Valori}}]{2007A&A...476..349F}
{Fuhrmann}, M., {Seehafer}, N., \& {Valori}, G. 2007, \aap, 476, 349

\bibitem[{{Galsgaard} {et~al.}(2003){Galsgaard}, {Titov}, \&
  {Neukirch}}]{2003ApJ...595..506G}
{Galsgaard}, K., {Titov}, V.~S., \& {Neukirch}, T. 2003, \apj, 595, 506

\bibitem[{{Gary}(2001)}]{2001SoPh..203...71G}
{Gary}, G.~A. 2001, \solphys, 203, 71

\bibitem[{{Gibson} {et~al.}(2004){Gibson}, {Fan}, {Mandrini}, {Fisher}, \&
  {Demoulin}}]{2004ApJ...617..600G}
{Gibson}, S.~E., {Fan}, Y., {Mandrini}, C., {Fisher}, G., \& {Demoulin}, P.
  2004, \apj, 617, 600

\bibitem[{{Green} \& {Kliem}(2009)}]{2009ApJ...700L..83G}
{Green}, L.~M. \& {Kliem}, B. 2009, \apjl, 700, L83

\bibitem[{{Guo} {et~al.}(2010){Guo}, {Schmieder}, {D{\'e}moulin}, {Wiegelmann},
  {Aulanier}, {T{\"o}r{\"o}k}, \& {Bommier}}]{Guo&al2010}
{Guo}, Y., {Schmieder}, B., {D{\'e}moulin}, P., {et~al.} 2010, \apj, accepted

\bibitem[{{Isenberg} \& {Forbes}(2007)}]{2007ApJ...670.1453I}
{Isenberg}, P.~A. \& {Forbes}, T.~G. 2007, \apj, 670, 1453

\bibitem[{{Keppens} {et~al.}(2003){Keppens}, {Nool}, {T\'oth}, \&
  {Goedbloed}}]{Keppensetal2003}
{Keppens}, R., {Nool}, M., {T\'oth}, G., \& {Goedbloed}, J. 2003, Comp. Phys.
  Comm., 153, 317
\bibitem[{{Kliem} {et~al.}(2004){Kliem}, {Titov}, \&
  {T{\"o}r{\"o}k}}]{2004A&A...413L..23K}
{Kliem}, B., {Titov}, V.~S., \& {T{\"o}r{\"o}k}, T. 2004, \aap, 413, L23

\bibitem[{{Kliem} \& {T{\"o}r{\"o}k}(2006)}]{2006PhRvL..96y5002K}
{Kliem}, B. \& {T{\"o}r{\"o}k}, T. 2006, Phys.~Rev.~Lett., 96, 255002

\bibitem[{{Leka} {et~al.}(1996){Leka}, {Canfield}, {McClymont}, \& {van
  Driel-Gesztelyi}}]{1996ApJ...462..547L}
{Leka}, K.~D., {Canfield}, R.~C., {McClymont}, A.~N., \& {van Driel-Gesztelyi},
  L. 1996, \apj, 462, 547

\bibitem[{{Leka} {et~al.}(2005){Leka}, {Fan}, \&
  {Barnes}}]{2005ApJ...626.1091L}
{Leka}, K.~D., {Fan}, Y., \& {Barnes}, G. 2005, \apj, 626, 1091

\bibitem[{{Lites} {et~al.}(1995){Lites}, {Low}, {Martinez Pillet}, {Seagraves},
  {Skumanich}, {Frank}, {Shine}, \& {Tsuneta}}]{1995ApJ...446..877L}
{Lites}, B.~W., {Low}, B.~C., {Martinez Pillet}, V., {et~al.} 1995, \apj, 446,
  877

\bibitem[{{Low} \& {Lou}(1990)}]{1990ApJ...352..343L}
{Low}, B.~C. \& {Lou}, Y.~Q. 1990, \apj, 352, 343

\bibitem[{{McClymont} {et~al.}(1997){McClymont}, {Jiao}, \&
  {Mikic}}]{1997SoPh..174..191M}
{McClymont}, A.~N., {Jiao}, L., \& {Mikic}, Z. 1997, \solphys, 174, 191

\bibitem[{{McKenzie} \& {Canfield}(2008)}]{2008A&A...481L..65M}
{McKenzie}, D.~E. \& {Canfield}, R.~C. 2008, \aap, 481, L65

\bibitem[{{Metcalf} {et~al.}(2008){Metcalf}, {Derosa}, {Schrijver}, {Barnes},
  {van Ballegooijen}, {Wiegelmann}, {Wheatland}, {Valori}, \&
  {McTtiernan}}]{2008SoPh..247..269M}
{Metcalf}, T.~R., {Derosa}, M.~L., {Schrijver}, C.~J., {et~al.} 2008, \solphys,
  247, 269

\bibitem[{{Metcalf} {et~al.}(1995){Metcalf}, {Jiao}, {McClymont}, {Canfield},
  \& {Uitenbroek}}]{1995ApJ...439..474M}
{Metcalf}, T.~R., {Jiao}, L., {McClymont}, A.~N., {Canfield}, R.~C., \&
  {Uitenbroek}, H. 1995, \apj, 439, 474

\bibitem[{{Moon} {et~al.}(2002){Moon}, {Choe}, {Yun}, {Park}, \&
  {Mickey}}]{2002ApJ...568..422M}
{Moon}, Y., {Choe}, G.~S., {Yun}, H.~S., {Park}, Y.~D., \& {Mickey}, D.~L.
  2002, \apj, 568, 422

\bibitem[{{Nakagawa} \& {Raadu}(1972)}]{1972SoPh...25..127N}
{Nakagawa}, Y. \& {Raadu}, M.~A. 1972, \solphys, 25, 127

\bibitem[{{Pevtsov} {et~al.}(1995){Pevtsov}, {Canfield}, \&
  {Metcalf}}]{1995ApJ...440L.109P}
{Pevtsov}, A.~A., {Canfield}, R.~C., \& {Metcalf}, T.~R. 1995, \apjl, 440, L109

\bibitem[{{Priest} \& {D{\'e}moulin}(1995)}]{1995JGR...10023443P}
{Priest}, E.~R. \& {D{\'e}moulin}, P. 1995, \jgr, 100, 23443

\bibitem[{{R{\' e}gnier} {et~al.}(2002){R{\' e}gnier}, {Amari}, \& {Kersal{\'
  e}}}]{2002A&A...392.1119R}
{R{\' e}gnier}, S., {Amari}, T., \& {Kersal{\' e}}, E. 2002, \aap, 392, 1119

\bibitem[{{R{\'e}gnier} \& {Amari}(2004)}]{2004A&A...425..345R}
{R{\'e}gnier}, S. \& {Amari}, T. 2004, \aap, 425, 345

\bibitem[{{Roumeliotis}(1996)}]{1996ApJ...473.1095R}
{Roumeliotis}, G. 1996, \apj, 473, 1095

\bibitem[{{Roussev} {et~al.}(2003){Roussev}, {Forbes}, {Gombosi}, {Sokolov},
  {DeZeeuw}, \& {Birn}}]{2003ApJ...588L..45R}
{Roussev}, I.~I., {Forbes}, T.~G., {Gombosi}, T.~I., {et~al.} 2003, \apjl, 588,
  L45

\bibitem[{{Sakurai}(1981)}]{1981SoPh...69..343S}
{Sakurai}, T. 1981, \solphys, 69, 343

\bibitem[{{Sakurai}(1989)}]{1989SSRv...51...11S}
{Sakurai}, T. 1989, Space Sci. Rev., 51, 11

\bibitem[{{Schmidt}(1964)}]{1964NASSP..50..107S}
{Schmidt}, H.~U. 1964, NASA Special Publication, 50, 107

\bibitem[{{Schrijver} {et~al.}(2008{\natexlab{a}}){Schrijver}, {DeRosa},
  {Metcalf}, {Barnes}, {Lites}, {Tarbell}, {McTiernan}, {Valori}, {Wiegelmann},
  {Wheatland}, {Amari}, {Aulanier}, {D{\'e}moulin}, {Fuhrmann}, {Kusano},
  {R{\'e}gnier}, \& {Thalmann}}]{2008ApJ...675.1637S}
{Schrijver}, C.~J., {DeRosa}, M.~L., {Metcalf}, T., {et~al.}
  2008{\natexlab{a}}, \apj, 675, 1637

\bibitem[{{Schrijver} {et~al.}(2006){Schrijver}, {DeRosa}, {Metcalf}, {Liu},
  {McTiernan}, {R{\'e}gnier}, {Valori}, {Wheatland}, \&
  {Wiegelmann}}]{2006SoPh..235..161S}
{Schrijver}, C.~J., {DeRosa}, M.~L., {Metcalf}, T.~R., {et~al.} 2006, \solphys,
  235, 161

\bibitem[{{Schrijver} {et~al.}(2008{\natexlab{b}}){Schrijver}, {Elmore},
  {Kliem}, {T{\"o}r{\"o}k}, \& {Title}}]{2008ApJ...674..586S}
{Schrijver}, C.~J., {Elmore}, C., {Kliem}, B., {T{\"o}r{\"o}k}, T., \& {Title},
  A.~M. 2008{\natexlab{b}}, \apj, 674, 586

\bibitem[{{Seehafer}(1978)}]{1978SoPh...58..215S}
{Seehafer}, N. 1978, \solphys, 58, 215

\bibitem[{{Seehafer}(1986)}]{1986SoPh..105..223S}
{Seehafer}, N. 1986, \solphys, 105, 223

\bibitem[{{Shafranov}(1966)}]{1966RvPP....2..103S}
{Shafranov}, V.~D. 1966, Reviews of Plasma Physics, 2, 103

\bibitem[{{Titov}(2007)}]{Titov2007}
{Titov}, V.~S. 2007, in Reconnection of magnetic fields : magnetohydrodynamics
  and collisionless theory and observations, ed. {Birn, J.~\& Priest, E.~R.}
  (Cambridge, UK: Cambridge University Press), 250--257

\bibitem[{{Titov} \& {D{\'e}moulin}(1999)}]{1999A&A...351..707T}
{Titov}, V.~S. \& {D{\'e}moulin}, P. 1999, \aap, 351, 707

\bibitem[{{Titov} {et~al.}(1999){Titov}, {D{\'e}moulin}, \&
  {Hornig}}]{1999ESASP.448..715T}
{Titov}, V.~S., {D{\'e}moulin}, P., \& {Hornig}, G. 1999, in ESA SP, Vol. 448,
  Magnetic Fields and Solar Processes, ed. {A.~Wilson et al.}, 715--722

\bibitem[{{Titov} {et~al.}(2003){Titov}, {Galsgaard}, \&
  {Neukirch}}]{2003ApJ...582.1172T}
{Titov}, V.~S., {Galsgaard}, K., \& {Neukirch}, T. 2003, \apj, 582, 1172

\bibitem[{{Titov} {et~al.}(2002){Titov}, {Hornig}, \&
  {D{\'e}moulin}}]{2002JGRA..107.1164T}
{Titov}, V.~S., {Hornig}, G., \& {D{\'e}moulin}, P. 2002, Journal of
  Geophysical Research (Space Physics), 107, 1164

\bibitem[{{Titov} {et~al.}(1993){Titov}, {Priest}, \&
  {Demoulin}}]{1993A&A...276..564T}
{Titov}, V.~S., {Priest}, E.~R., \& {Demoulin}, P. 1993, \aap, 276, 564

\bibitem[{{T{\"o}r{\"o}k} \& {Kliem}(2003)}]{2003A&A...406.1043T}
{T{\"o}r{\"o}k}, T. \& {Kliem}, B. 2003, \aap, 406, 1043

\bibitem[{{T{\"o}r{\"o}k} \& {Kliem}(2005)}]{2005ApJ...630L..97T}
{T{\"o}r{\"o}k}, T. \& {Kliem}, B. 2005, \apjl, 630, L97

\bibitem[{{T{\"o}r{\"o}k} \& {Kliem}(2007)}]{2007AN....328..743T}
{T{\"o}r{\"o}k}, T. \& {Kliem}, B. 2007, Astron.~Nachr., 328, 743

\bibitem[{{T{\"o}r{\"o}k} {et~al.}(2004){T{\"o}r{\"o}k}, {Kliem}, \&
  {Titov}}]{2004A&A...413L..27T}
{T{\"o}r{\"o}k}, T., {Kliem}, B., \& {Titov}, V.~S. 2004, \aap, 413, L27

\bibitem[{{Valori} {et~al.}(2007){Valori}, {Kliem}, \&
  {Fuhrmann}}]{2007SoPh..245..263V}
{Valori}, G., {Kliem}, B., \& {Fuhrmann}, M. 2007, \solphys, 245, 263

\bibitem[{{Valori} {et~al.}(2005){Valori}, {Kliem}, \&
  {Keppens}}]{2005A&A...433..335V}
{Valori}, G., {Kliem}, B., \& {Keppens}, R. 2005, \aap, 433, 335

\bibitem[{{van Ballegooijen}(2004)}]{2004ApJ...612..519V}
{van Ballegooijen}, A.~A. 2004, \apj, 612, 519

\bibitem[{{van Ballegooijen} {et~al.}(2007){van Ballegooijen}, {DeLuca},
  {Squires}, \& {Mackay}}]{ballegooijen07}
{van Ballegooijen}, A.~A., {DeLuca}, E.~E., {Squires}, K., \& {Mackay}, D.~H.
  2007, JASTP, 69, 24

\bibitem[{{Wheatland}(2000)}]{2000ApJ...532..616W}
{Wheatland}, M.~S. 2000, \apj, 532, 616

\bibitem[{{Wheatland} {et~al.}(2000){Wheatland}, {Sturrock}, \&
  {Roumeliotis}}]{2000ApJ...540.1150W}
{Wheatland}, M.~S., {Sturrock}, P.~A., \& {Roumeliotis}, G. 2000, \apj, 540,
  1150

\bibitem[{{Wiegelmann} {et~al.}(2006{\natexlab{a}}){Wiegelmann}, {Inhester},
  {Kliem}, {Valori}, \& {Neukirch}}]{2006A&A...453..737W}
{Wiegelmann}, T., {Inhester}, B., {Kliem}, B., {Valori}, G., \& {Neukirch}, T.
  2006{\natexlab{a}}, \aap, 453, 737

\bibitem[{{Wiegelmann} {et~al.}(2006{\natexlab{b}}){Wiegelmann}, {Inhester}, \&
  {Sakurai}}]{2006SoPh..233..215W}
{Wiegelmann}, T., {Inhester}, B., \& {Sakurai}, T. 2006{\natexlab{b}},
  \solphys, 233, 215

\bibitem[{{Wiegelmann} {et~al.}(2008){Wiegelmann}, {Thalmann}, {Schrijver},
  {Derosa}, \& {Metcalf}}]{2008SoPh..247..249W}
{Wiegelmann}, T., {Thalmann}, J.~K., {Schrijver}, C.~J., {Derosa}, M.~L., \&
  {Metcalf}, T.~R. 2008, \solphys, 247, 249

\bibitem[{{Yang} {et~al.}(1986){Yang}, {Sturrock}, \&
  {Antiochos}}]{1986ApJ...309..383Y}
{Yang}, W.~H., {Sturrock}, P.~A., \& {Antiochos}, S.~K. 1986, \apj, 309, 383

\bibitem[{{Zarro} {et~al.}(2000){Zarro}, {Canfield}, {Nitta}, {Myers},
  {Gregory}, {Qiu}, {Alexander}, {Hudson}, {Thompson}, \&
  {LaBonte}}]{2000SPD....31.0236Z}
{Zarro}, D.~M., {Canfield}, R.~C., {Nitta}, N., {et~al.} 2000, in Bulletin of
  the American Astronomical Society, Vol.~32, Bulletin of the American
  Astronomical Society, 817

\end{thebibliography}

\end{document}